\documentclass[sigconf, nonacm]{acmart}
\usepackage{tabularx}
\usepackage{multirow}

\usepackage{algorithm}
\usepackage{algpseudocode}

\usepackage{balance}

\AtBeginDocument{%
  }

\begin{document}

\title{Designing Human--GenAI Interaction for cMOOC Discussion Facilitation: Effects of a Collaborative AI-in-the-Loop Workflow on Social and Cognitive Presence}


\author{Jianjun Xiao}
\orcid{0000-0003-0000-9630}
\affiliation{%
  \institution{Research Center of Distance Education, Beijing Normal University}
  \city{Beijing}
  \country{China}}
\email{et\_shaw@126.com}

\author{Cixiao Wang}
\authornote{Corresponding author; Beijing Normal University Tang Scholar.}
\orcid{0000-0002-6053-8991}
\affiliation{%
  \institution{Research Center of Distance Education, Beijing Normal University}
  \city{Beijing}
  \country{China}}
\email{wangcixiao@bnu.edu.cn}


\renewcommand{\shortauthors}{Xiao et al.}
\renewcommand{\shorttitle}{Designing Human--GenAI Interaction for cMOOC Discussion Facilitation}

\begin{abstract}
Connectivist MOOCs (cMOOCs) rely on learner-driven interaction, yet their intentionally light facilitation makes it difficult to design generative AI participation that is both scalable and educationally productive. This design-based research study examined how human--GenAI interaction can be designed for discussion facilitation through a collaborative AI-in-the-loop workflow. Across two iterations in a five-week cMOOC ($N = 606$), we designed, deployed, and evaluated a facilitation system that combined network-structure-driven target selection, discourse-adaptive response roles, and mandatory human review before AI participation became visible in the community. Iteration~1 (Weeks~1--2) focused on refining the interaction design, showing that the most sustainable facilitation patterns were Guide (70.4\%) and Amplifier (28.5\%) responses and yielding explicit moderation standards for publishable AI participation. Iteration~2 (Weeks~3--5) examined how different forms of AI-mediated interaction related to social and cognitive presence. AI participation selectively enhanced Open Communication ($r = 0.188$, $p = .006$), Networked Cohesion ($r = 0.274$, $p < .001$), and overall social presence ($r = 0.162$, $p = .015$), while cognitive presence showed no overall improvement. More importantly, direct learner--agent interaction was associated with significantly higher social presence ($r = 0.186$, $p = .004$) and higher-order cognitive indicators---Integration ($r = 0.206$, $p = .001$) and Resolution/Creation ($r = 0.350$, $p < .001$)---than mere co-presence in AI-involved threads. The findings suggest that effective GenAI-supported discussion depends less on AI presence alone than on interaction design: reciprocal exchange, discourse-adaptive facilitation roles, and collaborative human review appear to be key conditions for productive AI participation in online learning communities.
\end{abstract}





\maketitle

\balance

\section{Introduction}

Massive Open Online Courses (MOOCs) have substantially broadened access to learning across geographic regions and learner populations, becoming an established modality within open and continuing higher education \citep{fidalgo-blanco_massive_2016, gao_diversity_2025}. However, sustaining meaningful engagement over time remains a persistent challenge. Research consistently demonstrates that social, cognitive, and teaching presence---the three pillars of the Community of Inquiry (CoI) framework \citep{garrison_critical_2001}---are strongly associated with participation quality, learner satisfaction, and perceived learning in online environments \citep{guo_community_2021, richardson_social_2017}. In large-scale MOOCs, these presences often remain underdeveloped due to limited instructor capacity and high learner attrition \citep{akyol_impact_2011, bai_more_2023, kovanovic_exploring_2018, poquet_social_2018}.

This challenge is particularly acute in connectivist MOOCs (cMOOCs), which emphasize learner autonomy, distributed knowledge production, and self-organized participation \citep{siemens_connectivism_2005, joksimovic_exploring_2018, chen_theoretical_2022}. Unlike content-centered xMOOCs, cMOOCs intentionally maintain minimal facilitation to preserve learner agency. This design choice, however, creates a practical dilemma: while peer interaction can be substantive, discourse often remains at surface levels without consistent facilitative support \citep{anderson_three_2011, bai_more_2023}. What is needed is a facilitation mechanism that is scalable, unobtrusive, and compatible with peer-driven discourse dynamics.

Pedagogical Conversational Agents (PCAs), enabled by recent advances in large language models (LLMs), offer a promising solution. PCAs can engage in dialogue, provide socio-emotional support, and scaffold reasoning processes \citep{johnson_pedagogical_2018, ortega-ochoa_empathic_2024, fuchs_exploring_2023}. In cMOOCs, a PCA designed as a learning partner rather than an authoritative tutor may address facilitation gaps while preserving the learner-driven nature of discourse. In the present study, the PCA served as the concrete form of GenAI participation. However, deploying LLM-driven PCAs in authentic educational settings raises legitimate concerns about content quality, role consistency, and the necessity for human oversight \citep{yan_promises_2024, kasneci_chatgpt_2023}. More fundamentally, the design problem is not simply whether GenAI should participate in discussion, but how human--GenAI interaction should be configured so that PCA participation becomes educationally useful rather than merely visible \citep{liu_improving_2024, ba_investigating_2025}.

Current research offers only partial guidance on this question. Most PCA studies examine structured courses or small-group settings \citep{han_faq_2022, songheehan_equity_2024, xing_unveiling_2026}, often treating GenAI support as a unitary intervention rather than distinguishing among interaction designs \citep{bastani_generative_2025}. Yet several design decisions are likely to matter in large-scale online discussion: \textit{when} GenAI should intervene, \textit{how} it should respond, \textit{whether} its participation should be filtered through human review, and \textit{what kind} of learner--agent interaction is educationally consequential. In community settings, mere PCA co-presence may not produce the same outcomes as reciprocal learner--agent exchange, and publishable PCA participation may require governance mechanisms that are integral to interaction design rather than secondary quality-control procedures.

To address these gaps, this study employed a design-based research (DBR) approach \citep{anderson_designbased_2012} to iteratively design, deploy, and evaluate a collaborative AI-in-the-loop facilitation workflow in a five-week cMOOC with 606 learners. The workflow, implemented through \textit{FaciHub}, combined network-structure-based focal post selection, emergent facilitation roles, and mandatory human review prior to publication. Iteration~1 focused on refining these interaction design decisions in authentic deployment. Iteration~2 examined how different forms of AI-mediated participation---with-PCA versus without-PCA discussion and, within intervention threads, direct interaction versus co-presence---were associated with social and cognitive presence.

This study makes three contributions. First, it reframes GenAI-supported facilitation in cMOOCs as an \textit{interaction design} problem and develops a collaborative AI-in-the-loop workflow grounded in networked discussion dynamics. Second, it provides deployment evidence showing which facilitation roles and moderation criteria made PCA participation publishable and contextually appropriate. Third, it offers empirical evidence that discussion outcomes depend less on PCA presence alone than on the form of learner--agent interaction, thereby yielding evidence-based design implications for GenAI participation in online learning communities.

\section{Theoretical Foundations}

\subsection{Community of Inquiry (CoI) Framework}

The CoI framework \citep{garrison_critical_1999} conceptualizes online learning through the interaction of three interdependent presences. \textit{Social presence} is ``the ability of participants in a CoI to project their characteristics into the community, thereby presenting themselves to other participants as `real people' '' \citep{garrison_critical_2001}. It encompasses affective expression, open communication, and group cohesion---dimensions that establish relational trust and psychological safety as prerequisites for sustained dialogue. \textit{Cognitive presence} refers to ``the extent to which the participants in any particular configuration of a community of inquiry can construct meaning through sustained dialogue'' \citep{garrison_critical_1999}. It unfolds through a four-phase practical inquiry cycle: triggering event, exploration, integration, and resolution \citep{garrison_critical_2001}. \textit{Teaching presence} includes design and organization, facilitation of discourse, and direct instruction. Rather than functioning solely as content delivery, teaching presence structures participation and helps discussion advance from idea exchange to deeper inquiry \citep{shea_reexamination_2010}.

These three presences are mutually reinforcing: social presence maintains open exchange, cognitive presence reflects the depth of collective meaning-making, and teaching presence provides direction and coherence \citep{kozan_interrelationships_2014, akyol_impact_2011}. A meta-analysis by \citet{richardson_social_2017} demonstrates robust links between social presence and learner satisfaction. Teaching presence---through task design \citep{rolim_networkbased_2019}, support strategies \citep{ba_investigating_2025}, platform affordances \citep{kong_exploratory_2025}, and interaction structure \citep{wang_what_2023, weidlich_extending_2023}---can shape presence trajectories.

\subsection{CoI in cMOOCs: Key Differences}

xMOOCs and cMOOCs differ in instructional design and in how teaching presence is enacted, which in turn shapes social and cognitive presence \citep{anderson_three_2011}. xMOOCs typically follow a structured, content-centered model with instructor-led delivery, predetermined learning paths, and formal assessment \citep{knox_learning_2018}. This structure provides clear teaching presence, but social and cognitive presence may remain peripheral when peer interaction is weakly integrated. In contrast, cMOOCs are grounded in connectivist principles \citep{siemens_connectivism_2005, joksimovic_exploring_2018}. They emphasize distributed content creation, peer-assisted learning, and minimal instructor intervention. Knowledge is co-constructed through networked interaction, and teaching presence is intentionally light to preserve learner agency \citep{chen_theoretical_2022, chen_new_2025, anderson_three_2011}. Table~\ref{tab:coi_comparison} summarizes how the three presences manifest differently across the two MOOC types.

\begin{table*}
\caption{Comparison of CoI presences in xMOOCs and cMOOCs}
\label{tab:coi_comparison}
\begin{tabular}{p{0.15\textwidth}p{0.38\textwidth}p{0.38\textwidth}}
\toprule
\textbf{Presence} & \textbf{xMOOC} & \textbf{cMOOC} \\
\midrule
Cognitive & Individual and task-focused; structured progression through predefined content & Networked and collaborative; develops through emergent connections and distributed knowledge creation \\
Social & Often peripheral, with weaker peer exchange & Central to participation, with community-oriented interaction and peer linkage \\
Teaching & Explicit instructional control and structured guidance & Distributed facilitation, peer teaching, and co-constructed direction \\
\bottomrule
\end{tabular}
\end{table*}

MOOC-specific validation studies suggest that CoI presences operate differently at scale. Large-sample analyses have supported the original three-factor structure but also reported improved fit for an expanded six-factor model, in which design and organization, affective expression, and resolution emerge as distinct dimensions \citep{kovanovic_exploring_2018, bai_more_2023}. These findings indicate that scale, participation diversity, and reduced facilitation visibility render certain presence dimensions more salient than in smaller online courses.

Although cMOOCs can stimulate active peer discourse, facilitation quality is often uneven. Existing cMOOC research has examined content production \citep{xu_research_2024, bai_impact_2023}, social capital \citep{li_exploring_2022, joksimovic_exploring_2018}, and network interaction patterns \citep{li_dynamics_2025, wang_interaction_2017, tian_measurement_2026}, yet CoI-oriented analysis of social and cognitive presence remains limited. Two questions are therefore central to this study: how can facilitative support be provided when teaching presence is distributed, and how should presence be measured to fit cMOOC interaction patterns? The following subsections address these questions.

\subsection{Implications for the Current Study}

\subsubsection{Distributed teaching presence and alternative facilitation}

In cMOOCs, teaching presence is spread across the community rather than held by a single instructor \citep{anderson_three_2011}. Facilitation emerges from peer contributions and collective sense-making. However, this distributed model often results in uneven and inconsistent support. Many learners share content, but few take on facilitative roles such as guiding discussion, asking follow-up questions, or synthesizing ideas \citep{poquet_social_2018, wang_interaction_2017}. Without stable facilitation, discussions may remain shallow or stall before reaching deeper inquiry \citep{zhang_comparative_2023}.

This gap points to the need for a lightweight, scalable mechanism that can provide consistent facilitative support without replacing learner agency. Prior work has shown that instructional interventions can be evaluated through their effects on social and cognitive presence \citep{rolim_networkbased_2019}. Following this approach, a PCA can serve as one way to operationalize teaching presence in cMOOCs. The goal is not to replace peer discourse but to enact facilitative functions that stabilize and extend it.

Therefore, this study designs \textit{FaciHub} as a form of distributed teaching presence. By taking on multiple facilitative roles within networked discussions, \textit{FaciHub} aims to support social and cognitive presence while preserving the learner-driven nature of cMOOCs.

\subsubsection{Adapting social and cognitive presence for cMOOCs}\label{sec:coi_adaptations}

The original CoI coding schemes assume a small, stable learning group in which members interact repeatedly and progressively build shared identity \citep{garrison_critical_2001, anderson_three_2011}. This assumption underpins both the Group Cohesion category---which presupposes bounded membership and interpersonal bonding---and the Resolution category---which conceptualizes inquiry closure as problem application within a task-defined framework. cMOOCs violate both assumptions: participants form weak, transient ties; engage with different peers across threads; and may never interact with the same person twice \citep{granovetter_strength_1973, joksimovic_exploring_2018}. Furthermore, in professional learning communities, the endpoint of inquiry frequently involves the creation of original knowledge artifacts rather than task resolution alone \citep{chen_new_2025, chen_theoretical_2022}. Two adaptations are therefore necessary---not cosmetic terminological adjustments, but theoretically motivated reconceptualizations fitted to the structural and functional conditions of cMOOC learning.

\paragraph{From Group Cohesion to Networked Cohesion.}

Group cohesion in the original CoI framework refers to the interpersonal bonds that form within a closed, stable group: the shared sense of community that motivates sustained commitment to dialogue \citep{garrison_critical_2001}. This conception is grounded in small-group psychology, where cohesion accrues through repeated interaction, mutual familiarity, and deepening affective trust. In cMOOCs, no such stable group exists. Participation is open, continuous, and structurally fluid; learners navigate a distributed network of peers rather than a fixed cohort. Social network theory provides an alternative framing: in networks characterized by weak ties and open participation, community belonging is expressed not through dyadic closeness but through shared language, collective norms, and identification with the professional community as a whole---markers that signal membership in a broader learning network rather than a bounded group \citep{granovetter_strength_1973, siemens_connectivism_2005}. This reconceptualization is supported empirically: large-sample CoI validation studies in MOOCs report that cohesion dimensions operate differently at scale, with network-level belonging emerging as structurally distinct from interpersonal affiliation \citep{kovanovic_exploring_2018, bai_more_2023}. Networked learning research further confirms that community identification---rather than dyadic bonding---is the characteristic sociality mechanism in cMOOCs, manifested in learners' use of community-specific terminology, shared referential frames, and collective professional language \citep{joksimovic_exploring_2018, poquet_social_2018}. Accordingly, \textit{Group Cohesion} is reconceptualized as \textit{Networked Cohesion} (NC), with coding indicators reoriented toward network-level expressions of belonging.

\paragraph{From Resolution to Resolution and Creation.}

In the original practical inquiry model, resolution denotes the application of newly constructed understanding to the problem that triggered the inquiry cycle \citep{garrison_critical_2001}. While this captures an important endpoint for academic task contexts, it does not adequately represent a theoretically and empirically prominent outcome in professional learning communities: the production of original knowledge artifacts. Knowledge creation scholarship distinguishes between knowledge \textit{use} and knowledge \textit{production}: professional learning communities generate new professional tools---lesson designs, teaching frameworks, assessment instruments---that externalize tacit expertise and contribute to a shared knowledge base \citep{paavola_models_2004}. The trialogical model of learning \citep{paavola_models_2004} explicitly identifies artifact creation as a third mode of knowing, analytically distinct from knowledge transmission and from social construction through dialogue, and particularly salient in expert communities engaged in sustained professional practice. In teacher professional development specifically, this dimension manifests as practitioners' production of shareable professional knowledge that advances both individual competence and community capability \citep{stoll_professional_2006, voogt_teacher_2011}. Empirical work in cMOOC contexts confirms that artifact creation constitutes an observable, analytically distinct endpoint of cognitive engagement that the original resolution category cannot adequately capture \citep{xu_research_2024, duan_connectivist_2025}. Accordingly, \textit{Resolution} is expanded to \textit{Resolution and Creation} (RC), encompassing both knowledge application (RC-1: Applying Solutions) and original artifact production (RC-2: Artifact Creation) as distinct higher-order cognitive indicators \citep{wang_interaction_2017}.

These adaptations are operationalized in the Iteration~2 coding framework (Section~\ref{sec:iteration2_coding_framework}).

\section{Related Work}

This section reviews empirical work relevant to the design and evaluation of collaborative GenAI-supported facilitation. Two themes structure the review: (1) how PCAs can operationalize teaching presence in cMOOCs through intervention targeting, facilitation roles, and human oversight; and (2) why different forms of learner--PCA interaction may produce different social and cognitive outcomes.

\subsection{PCAs as Operationalization of Teaching Presence}

\subsubsection{Network-Structure-Driven Intervention in Online Learning}

Network analysis has been widely employed to understand interaction patterns in online learning communities \citep{rolim_networkbased_2019, zou_exploring_2021}. Centrality metrics can identify influential posts, active participants, or emerging discussion clusters \citep{li_exploring_2022}. In MOOCs, network-based approaches have been applied to analyze learner engagement, knowledge flow, and community structure \citep{joksimovic_exploring_2018}.

However, most studies employ network analysis for post-hoc description rather than real-time intervention \citep{xiao_exploring_2026}. Few have explored how network structure can guide where and when a PCA should participate \citep{feng_group_2025}. In cMOOCs, where discussions are decentralized and non-linear, targeting high-centrality posts may help PCA responses reach more learners while aligning with the emergent nature of community discourse \citep{battiston_higherorder_2025}. This study employs a hypergraph-based approach to identify focal posts for PCA intervention based on closeness centrality, which captures a node's average distance to all other nodes and thus indicates potential reach within the discussion network.

\subsubsection{Facilitator Roles in Online Discussions}

Research on online facilitation identifies several key roles that support productive discussion: guiding conversation, amplifying valuable contributions, providing socio-emotional support, and prompting critical reflection \citep{shaikh_role_2012, zhang_comparative_2023, bai_how_2025}. In instructor-led courses, a designated facilitator typically performs these roles. In cMOOCs, such roles must emerge from peer contributions or platform-mediated features.

Recent research on conversational agent design has explored how different agent roles influence learning outcomes. \citet{nguyen_role_2023} demonstrated that role design affects how learners engage with agent-generated content. Studies on human facilitators indicate that when facilitative behaviors appear---such as summarizing ideas, posing follow-up questions, or affirming contributions---discussions tend to be more sustained and substantive \citep{zhang_comparative_2023}. Yet such behaviors are unevenly distributed in peer-driven communities \citep{anderson_three_2011, shea_building_2022}. This suggests that effective PCAs should enact multiple facilitative roles matched to different contribution types, rather than adopting a single instructor persona.

\subsubsection{Content Quality and Human Oversight in LLM-Driven Agents}

LLMs enable PCAs to generate fluent and context-responsive dialogue at scale. However, deployment in authentic educational settings introduces quality risks extending beyond linguistic fluency \citep{yan_promises_2024, kasneci_chatgpt_2023}. LLM-generated responses may hallucinate---fabricating facts, personal experiences, or instructional details absent from the discussion context. Responses may also drift from the assigned role, creating tone or intent mismatches with the pedagogical purpose of the interaction. These concerns are amplified in cMOOCs, where discussions are learner-generated, topically diverse, and rapidly evolving.

These risks raise a fundamental governance question: how should authority be allocated between AI systems and human professionals? Two paradigms are commonly distinguished: Human-in-the-Loop (HITL), where AI leads with human oversight, and AI-in-the-Loop (AITL), where human decision-making remains primary and AI provides computational support \citep{natarajan_humanintheloop_2025}. In education, AITL is particularly relevant because pedagogical decisions require professional judgment that current AI systems cannot reliably provide independently. Prior work on AI-assisted facilitation has predominantly adopted HITL designs \citep{ba_investigating_2025, song_interactions_2025}, but practical frameworks for human--AI collaborative review---including criteria, workflow, and decision rules---remain underdeveloped. This gap is not only a governance issue but also an interaction design issue: human review influences which PCA utterances become socially visible and therefore shapes the educational interaction that learners actually encounter.

\subsection{Effects of PCAs on Social and Cognitive Presence}

\subsubsection{Empirical Evidence from MOOC Contexts}

Empirical CoI studies in MOOCs indicate that cognitive presence tends to cluster in early inquiry phases (triggering and exploration), while higher-order phases (integration and resolution) occur less frequently \citep{liu_looking_2022, guo_community_2021}. Social presence is similarly uneven, with active participants demonstrating higher levels than peripheral participants \citep{poquet_social_2018, zou_exploring_2021}. Strengthening teaching presence is consistently identified as a key lever for improving both presences \citep{zou_exploring_2021, shea_reexamination_2010}, but how to provide such support in cMOOCs---where formal facilitation is intentionally minimal---remains an open question.

PCAs offer one potential approach. They can support social presence through self-disclosure prompts, empathic responses, and personalized addressing \citep{huang_facilitating_2025, huang_chatbots_2022, ortega-ochoa_empathic_2024}, and support cognitive presence through content-based scaffolding that encourages idea development and synthesis \citep{ba_investigating_2025, nguyen_role_2023, kong_exploratory_2025}. Recent research demonstrates that design choices---such as knowledge-oriented versus emotion-oriented scaffolding---produce differential effects across presence dimensions \citep{liu_improving_2024}, highlighting the need to examine how specific agent designs map to specific outcomes. However, most PCA studies are conducted in structured courses or small groups, with agents positioned as tutors or feedback tools \citep{han_faq_2022, songheehan_equity_2024}. Evidence from large-scale, learner-driven cMOOCs remains limited.

\subsubsection{Interaction Mode as a Moderating Factor}

Different learner--PCA interaction modes may shape presence in distinct ways. \citet{ba_investigating_2025} found that idea-oriented AI feedback enhanced cognitive presence but also triggered negative emotions, while task-oriented feedback promoted social interaction. Two modes are particularly relevant in cMOOC discussions: \textit{direct interaction}, where a learner exchanges replies with the agent, and \textit{co-presence}, where the agent participates in the same thread without direct exchange with a given learner. The distinction parallels research on social presence in computer-mediated communication, which differentiates between interacting with others and merely perceiving their presence \citep{prestridge_identifying_2024}. Research on human--AI group interaction suggests that agent positioning can shift interaction patterns \citep{feng_group_2025}, but comparative evidence on how these two modes affect social and cognitive presence in cMOOCs remains scarce. This distinction is central for evidence-based interaction design because it tests whether educational value derives from reciprocal exchange with the PCA or simply from the PCA's ambient participation in the discussion space.

\subsection{Research Questions}

The preceding review identifies two gaps. First, how to design scalable, discourse-adaptive human--GenAI interaction for cMOOC facilitation remains unclear, especially when human review is required before PCA participation becomes visible in the community. Second, whether different learner--PCA interaction modes (direct interaction vs.\ co-presence) produce differential presence outcomes has not been empirically examined. This study addresses both gaps through two research questions:

\begin{itemize}
    \item \textbf{RQ1 (Iteration~1):} How can a collaborative AI-in-the-loop interaction design for network-embedded facilitation be iteratively refined to support publishable and pedagogically appropriate PCA participation in cMOOCs?
    \item \textbf{RQ2 (Iteration~2):} How are different forms of PCA-mediated interaction---PCA participation and learner--PCA interaction mode (direct interaction vs.\ co-presence)---associated with social and cognitive presence in cMOOC discussions?
\end{itemize}

\section{Interaction Design of the Collaborative AI-in-the-Loop Workflow}

This section presents the initial interaction design of the collaborative AI-in-the-loop workflow, which was subsequently refined through the two DBR iterations. The workflow was designed to structure \textit{how} GenAI would participate in learner-driven discussion rather than merely adding a generic AI presence to the forum. It comprised three design decisions: (1) \textit{targeting design}, which determined which posts or comments would receive PCA attention; (2) \textit{response-role design}, which determined how the PCA would participate once a target was selected; and (3) \textit{participation governance design}, which determined how human facilitators would review PCA responses before they became publicly visible.

\begin{figure*}
    \centering
    \includegraphics[width=0.8\linewidth]{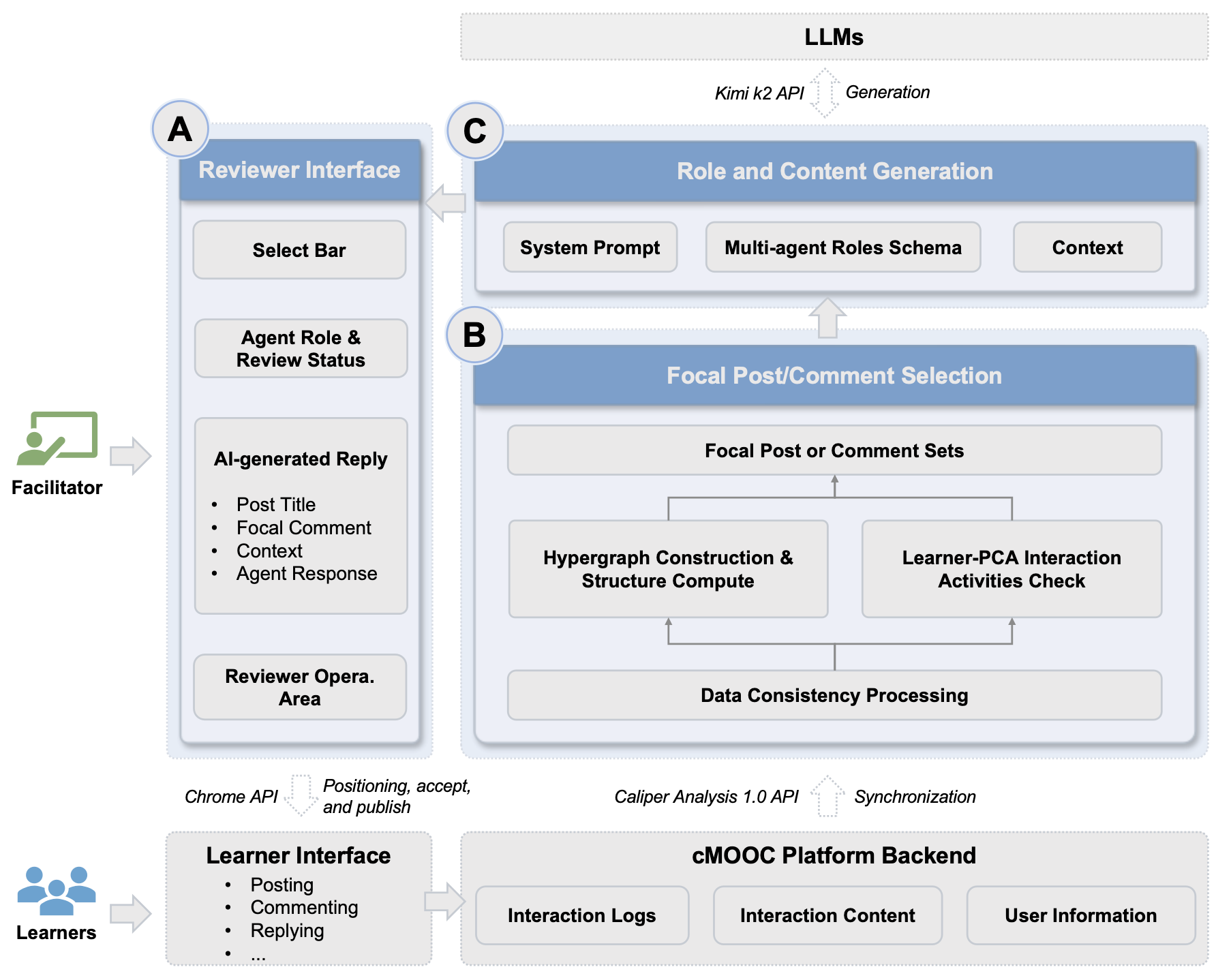}
    \caption{Workflow structure for PCA target selection, response generation, and human review}
    \label{fig:system_architecture}
\end{figure*}

To implement this workflow reliably in authentic cMOOC practice, and to make it reusable for facilitators, we developed \textit{FaciHub}---a Human--AI Collaboration Hub for Facilitators. As shown in Figure~\ref{fig:system_architecture}, \textit{FaciHub} operationalized the three design decisions through corresponding modules: the focal post/comment selection component (module B) supported the targeting design, the role-based content generation component (module C) supported the response-role design, and the facilitator review interface (module A) supported the participation governance design. \textit{FaciHub} is introduced here as the platform through which the collaborative AI-in-the-loop workflow could be enacted consistently and examined empirically.

\subsection{Participation Governance Design}\label{module:module_A}

\begin{figure*}
    \centering
    \includegraphics[width=1\linewidth]{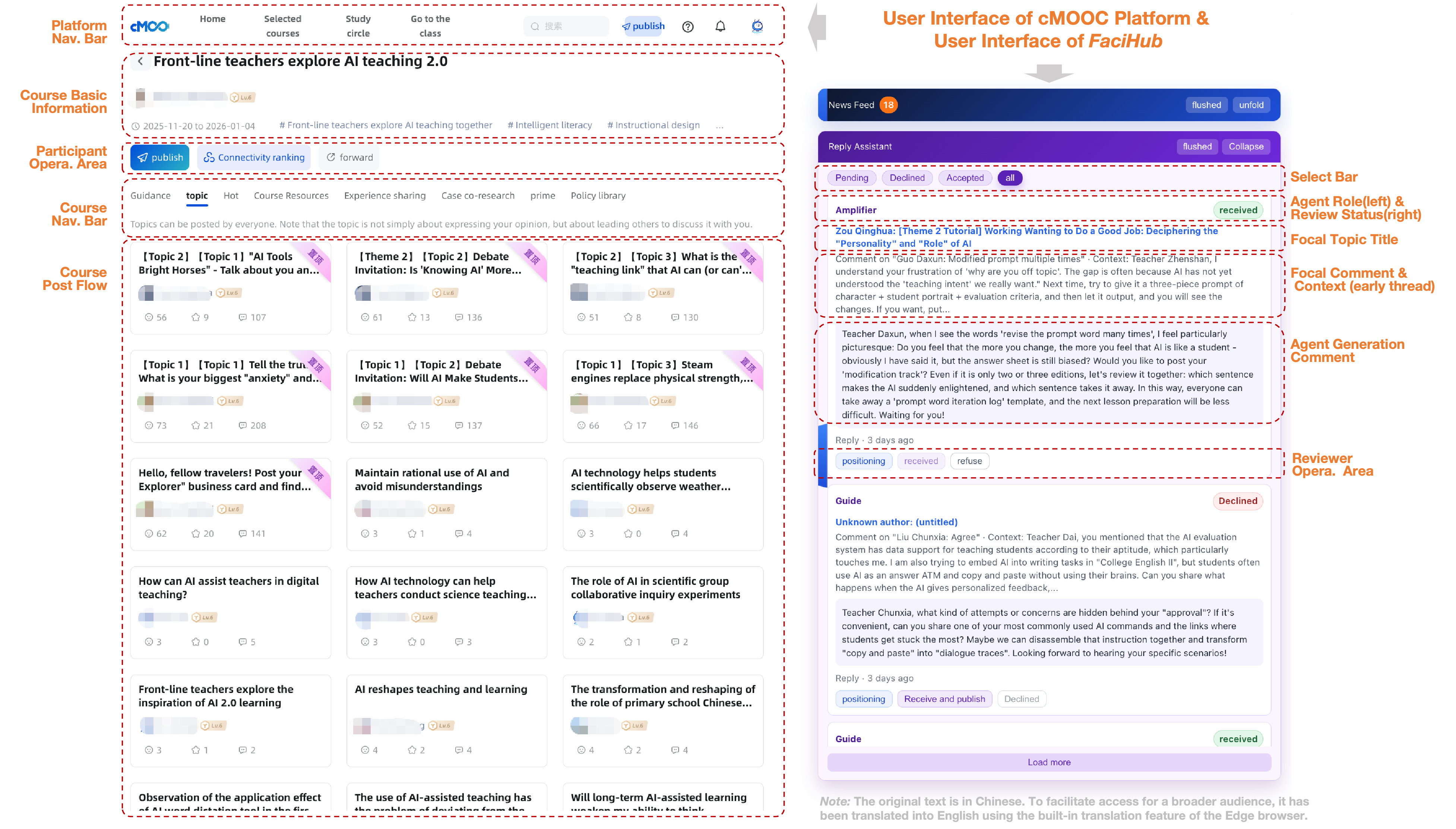}
    \caption{Interface for contextual inspection and human approval of PCA-generated replies}
    \label{fig:user_interface}
\end{figure*}

The workflow was integrated into the cMOOC platform via a Chrome extension that supported facilitator review and publication decisions. As shown in Figure~\ref{fig:user_interface}, facilitators could inspect candidate PCA replies together with the focal post, thread context, assigned role, and moderation status, then either approve or reject the response before publication. This design made human review a constitutive part of the learner-visible interaction: only approved PCA responses entered the discussion space.

This governance layer served two functions. Pedagogically, it ensured that PCA participation remained contextually appropriate and aligned with facilitation goals. Methodologically, it generated the acceptance and rejection records later used in Iteration~1 to refine moderation standards and prompt instructions. Thus, the interface was treated as the mechanism through which human judgment shaped publicly visible PCA participation.

\subsection{Targeting Design}\label{module:module_B}

The targeting design identified posts and comments warranting PCA responses through two daily mechanisms.

\begin{figure}
    \centering
    \includegraphics[width=1\linewidth]{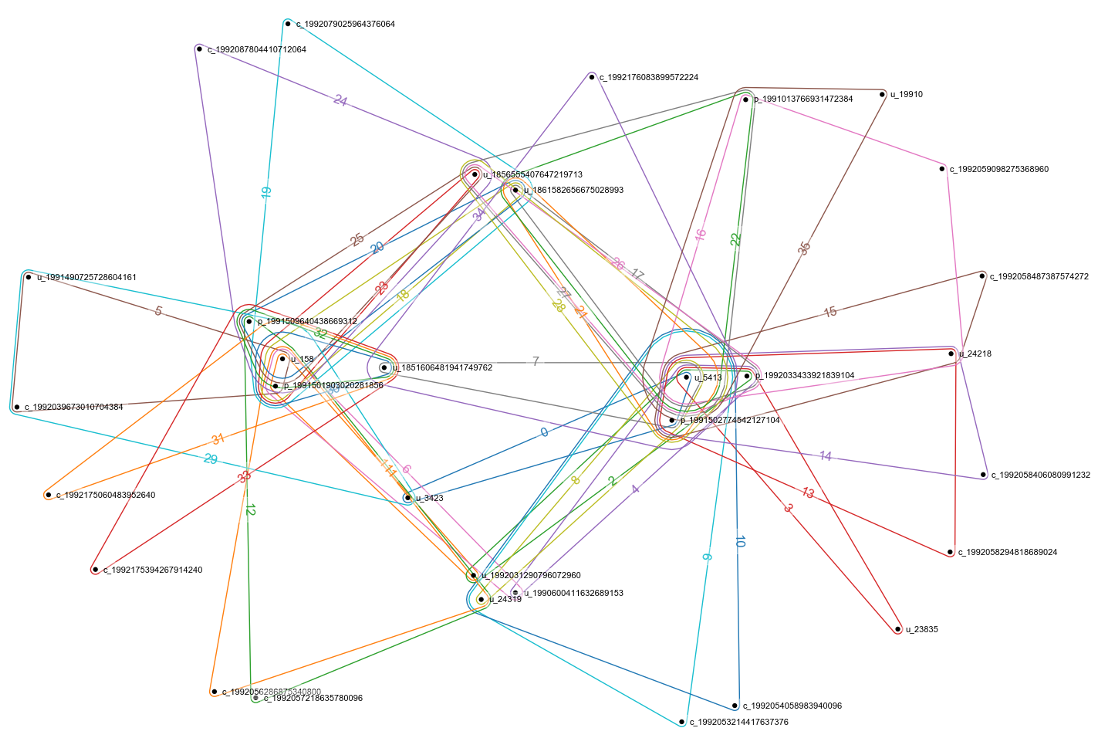}
    \caption{Example of hypergraph-based target selection for PCA participation}
    \label{fig:screening_mechanism}
\end{figure}

The first mechanism is \textbf{network-structure triggering}. Based on platform behavioral logs, the system constructs a hypergraph from the most recent two days of activity. In this representation, hyperedges represent interaction events, and nodes represent users (U), posts (P), and comments/replies (C)--that is "who interacts with whom and what the interaction entails." Figure~\ref{fig:screening_mechanism} illustrates an example network. The system ranks nodes by closeness centrality and selects the top 5\% of P-type and C-type nodes as candidate intervention targets. These targets represent active discussion areas, enabling PCA responses to reach more learners while respecting the emergent, non-predetermined pathways characteristic of cMOOCs.

The second mechanism is \textbf{learner-reply triggering}. When a learner directly replies to a PCA comment, that reply is automatically added as a target in the next scheduled processing run.

Together, these mechanisms linked PCA participation to network position and learner behavior rather than to a predetermined posting schedule. Candidate sets from both mechanisms were merged and used to construct the prompt context sent to the LLM (\textit{kimi-k2-turbo-preview}, $temperature=0.6$). Technical details are provided in Appendix~\ref{appendix:focal_post_selection}.

\subsection{Response-Role Design}\label{module:module_C}

The response-role design specified how the PCA should participate once a target had been selected. Four roles were initially configured to address different discourse conditions: Guide, Amplifier, Empathizer, and Critical Inquirer. The \textit{Guide} and \textit{Amplifier} roles draw on established facilitation practices in cMOOCs. The \textit{Empathizer} and \textit{Critical Inquirer} roles were included to address affective support and reflective questioning, respectively.

Each role was matched to target comments based on textual characteristics and followed a distinct interaction objective and linguistic style. The Guide role scaffolded deeper reflection and invited further discussion. The Amplifier role recognized the value of practice-based sharing and encouraged broader peer exchange. The Empathizer role provided emotional support and psychological safety. The Critical Inquirer role used gentle questioning to promote critical reflection and boundary awareness. In this sense, the role system functioned as an interaction logic rather than as a set of independent personas: it defined what kind of PCA participation would be attempted under different discourse conditions. Technical details on multi-agent role generation are provided in Appendix~\ref{appendix:multi_agent_role_generation}.

Taken together, these three design elements established the initial workflow examined in the DBR process. Iteration~1 evaluated whether the targeting, role-selection, and human-review arrangements produced publishable and pedagogically appropriate PCA participation. Iteration~2 then examined whether different forms of PCA-mediated interaction within this workflow were associated with different discussion outcomes.

\section{Research Design}

\subsection{Design-Based Research Approach}

\begin{figure*}
    \centering
    \includegraphics[width=0.8\linewidth]{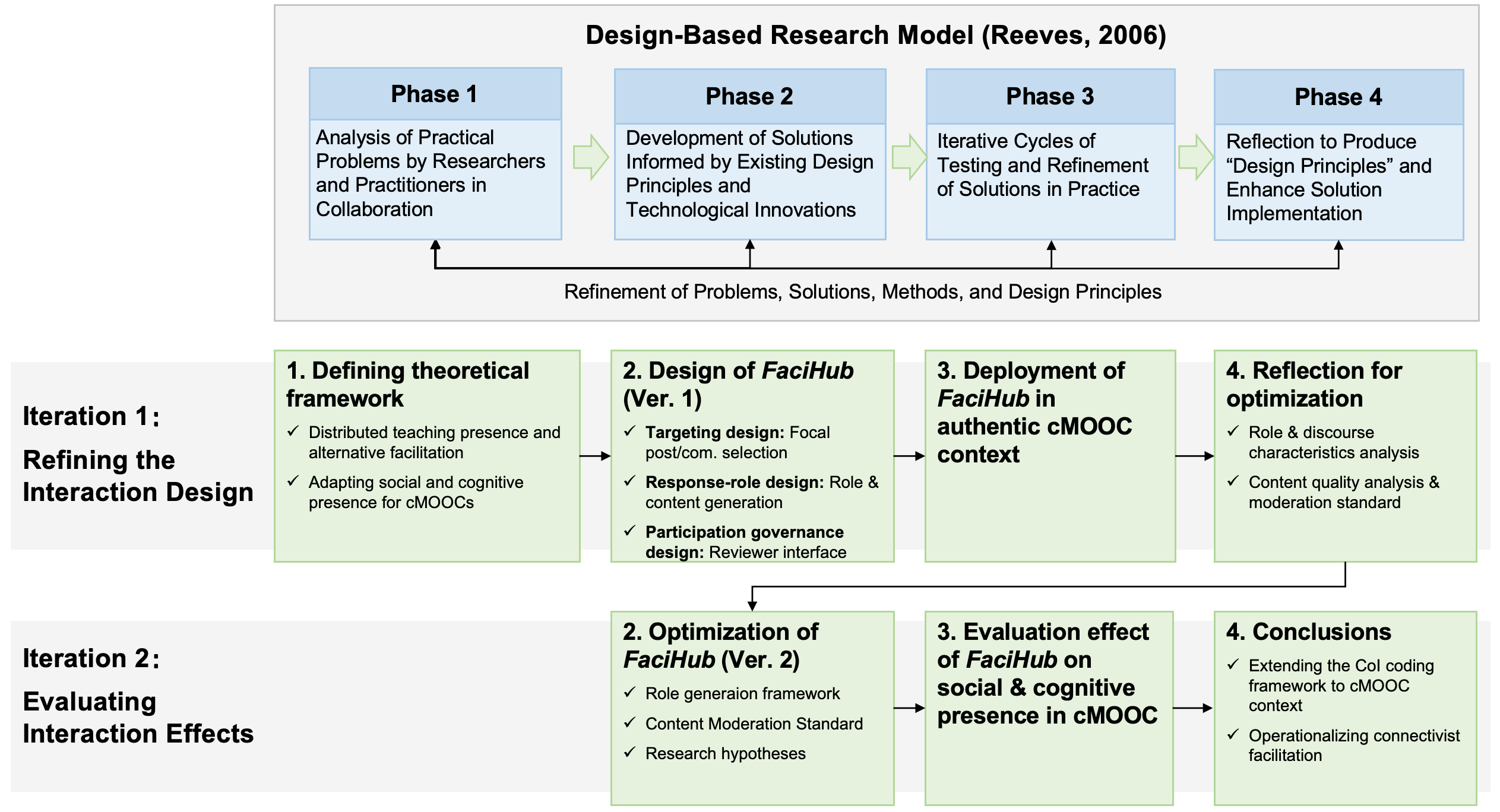}
    \caption{Overview of the DBR procedure aligned with Reeves' (2006) four-phase model}
    \label{fig:dbr_overview}
\end{figure*}

To address the two research questions, this study adopted a DBR approach \citep{reeves_design_2006} to iteratively design, deploy, and evaluate a collaborative AI-in-the-loop facilitation workflow in a five-week authentic cMOOC setting. Rather than treating GenAI participation as a single intervention, the DBR process focused on refining three interaction design decisions: (1) \textit{targeting design}---which discussion contributions should receive AI facilitation, (2) \textit{response-role design}---how the AI should participate in discourse, and (3) \textit{participation governance design}---how human review should regulate which AI responses become publicly visible. The study was organized into two iterations (Figure~\ref{fig:dbr_overview}):

\begin{itemize}
    \item \textbf{Iteration~1 (Weeks~1--2):} Addressing RQ1, this iteration deployed the workflow in authentic discussions, analyzed how discourse-adaptive roles and human review shaped publishable AI participation, and established moderation standards for AI-mediated interaction.
    \item \textbf{Iteration~2 (Weeks~3--5):} Addressing RQ2, this iteration examined how different forms of AI-mediated interaction---with-PCA versus without-PCA participation, and direct interaction versus co-presence---relate to social and cognitive presence, building on the moderation standards and role refinement from Iteration~1.
\end{itemize}

\subsection{Research Context}

The study was conducted in a Chinese cMOOC titled \textit{Front-line Teachers Explore AI Teaching 2.0} (November 24 to December 29, 2025), hosted on a university-developed open learning platform by a Chinese university. Designed with reference to UNESCO's \textit{AI Competency Framework for Teachers}, the course aimed to strengthen participants' understanding of AI in education and enhance their practical competencies. The five-week course comprised four thematic modules: (1) AI Attitude and Ethics, (2) AI Knowledge and Tools, (3) AI Teaching Methods, and (4) AI Professional Development. Participants engaged in discussions, experience sharing, and reflection on applying AI-supported pedagogical strategies in professional practice.

\begin{figure*}
    \centering
    \includegraphics[width=1\linewidth]{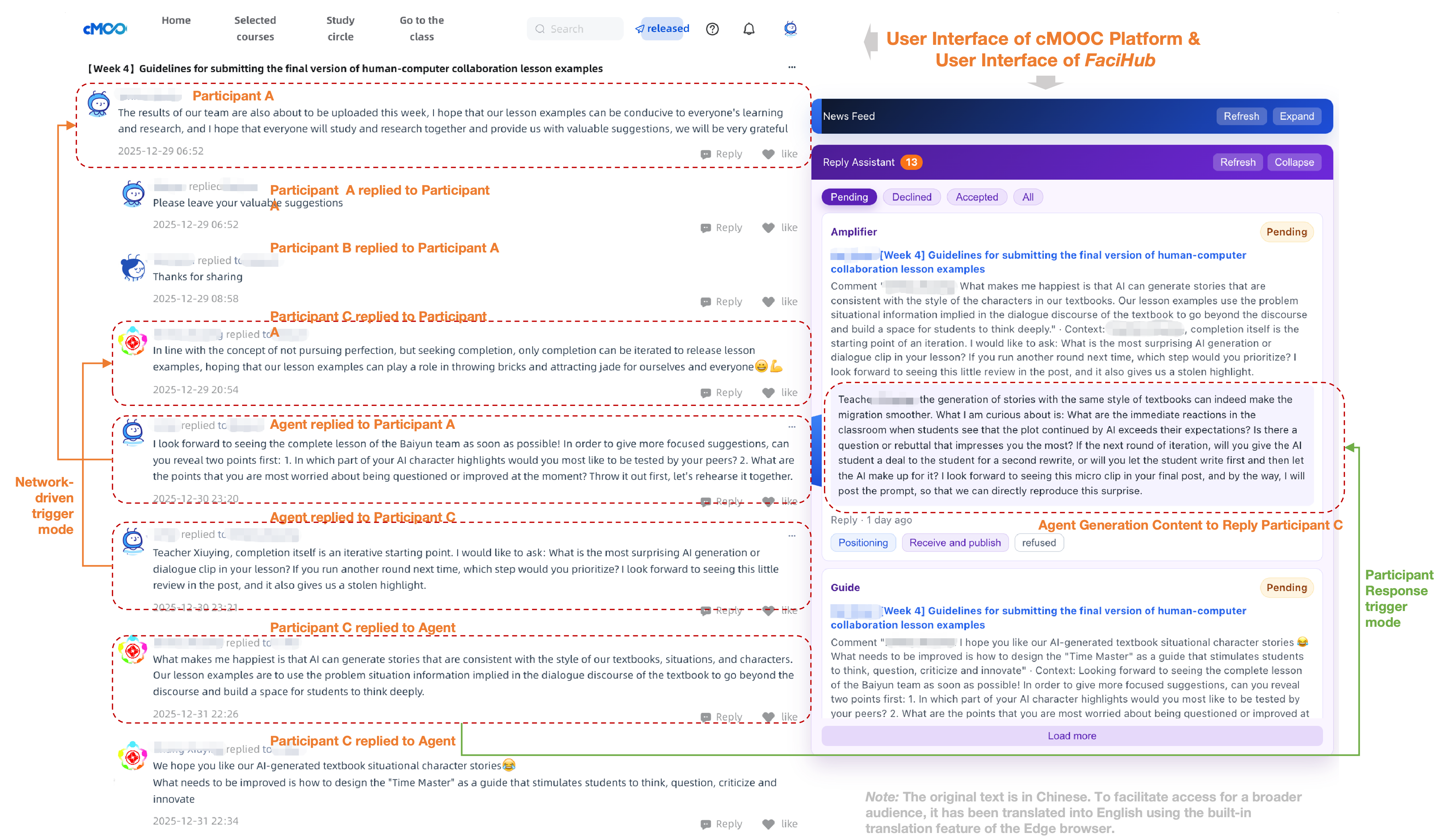}
    \caption{Example of a discussion thread with PCA participation}
    \label{fig:agent_participant_thread_example}
\end{figure*}

\textit{FaciHub} was embedded in discussion threads to augment facilitator--learner interaction. As illustrated in Figure~\ref{fig:agent_participant_thread_example}, the PCA participated alongside learners by generating context-aware replies to participant posts. All replies underwent human facilitator review prior to publication, making human review part of the interaction design rather than a post-hoc audit step. This arrangement positioned PCA contributions within natural peer discussion while ensuring that only pedagogically appropriate AI participation became visible in the community.

\subsection{Participants}

A total of 606 learners enrolled in the course. The majority were in-service teachers ($n = 468$, 77.2\%), followed by students ($n = 61$, 10.1\%), education administrators ($n = 26$, 4.3\%), industry practitioners ($n = 10$, 1.7\%), and others ($n = 11$, 1.8\%). Thirty participants (5.0\%) did not report their professional background. This composition reflected the course's focus on professional learning in AI-enhanced education, attracting participants primarily from educational practice and administration.

\section{Iteration~1: Refining the Interaction Design}

\subsection{Research Goals}

Iteration~1 addressed RQ1 by deploying the collaborative AI-in-the-loop facilitation workflow in the cMOOC during Weeks~1--2 and refining how GenAI should participate in discussion. This phase pursued three specific goals:

\begin{enumerate}
    \item To deploy the targeting and response-role design decisions in authentic community discussions.
    \item To analyze the distribution and content characteristics of PCA-generated responses across four emergent roles (Guide, Amplifier, Empathizer, Critical Inquirer), and to determine which roles should remain active based on observed discourse patterns.
    \item To evaluate generated responses through human facilitator review and refine moderation standards for publishable PCA participation, ensuring pedagogical appropriateness, contextual coherence, and role consistency.
\end{enumerate}

\subsection{Methodology}

\subsubsection{Procedure}

During Weeks~1--2, the facilitation workflow was deployed in the cMOOC forum. The targeting design identified candidate discussion targets daily based on the hypergraph representation of community interactions, and the response-role design generated replies through four emergent roles---Guide, Amplifier, Empathizer, and Critical Inquirer. All generated responses were routed to the facilitator review interface rather than being published directly.

Human facilitators reviewed each response and made accept/reject decisions. This participation-governance process served dual purposes: quality control and empirical grounding for the moderation standards that determined which PCA utterances entered the visible community discourse. Responses were evaluated along three predefined dimensions: role and task alignment, interactional appropriateness, and factual plausibility with contextual coherence. Although these dimensions were specified prior to deployment, the specific acceptance and rejection criteria were iteratively refined through facilitator--researcher discussion during the review process (see Section~\ref{sec:human_review_criteria}).

\subsubsection{Data Corpus}

During Weeks~1--2, \textit{FaciHub} generated 625 candidate responses for facilitator review. Of these, 446 responses (71.4\%) were accepted and published, while 179 (28.6\%) were rejected. Representative examples of learner posts and corresponding PCA responses for each role are provided in Appendix~\ref{appendix:role_examples}.

\begin{figure*}
    \centering
    \includegraphics[width=0.8\textwidth]{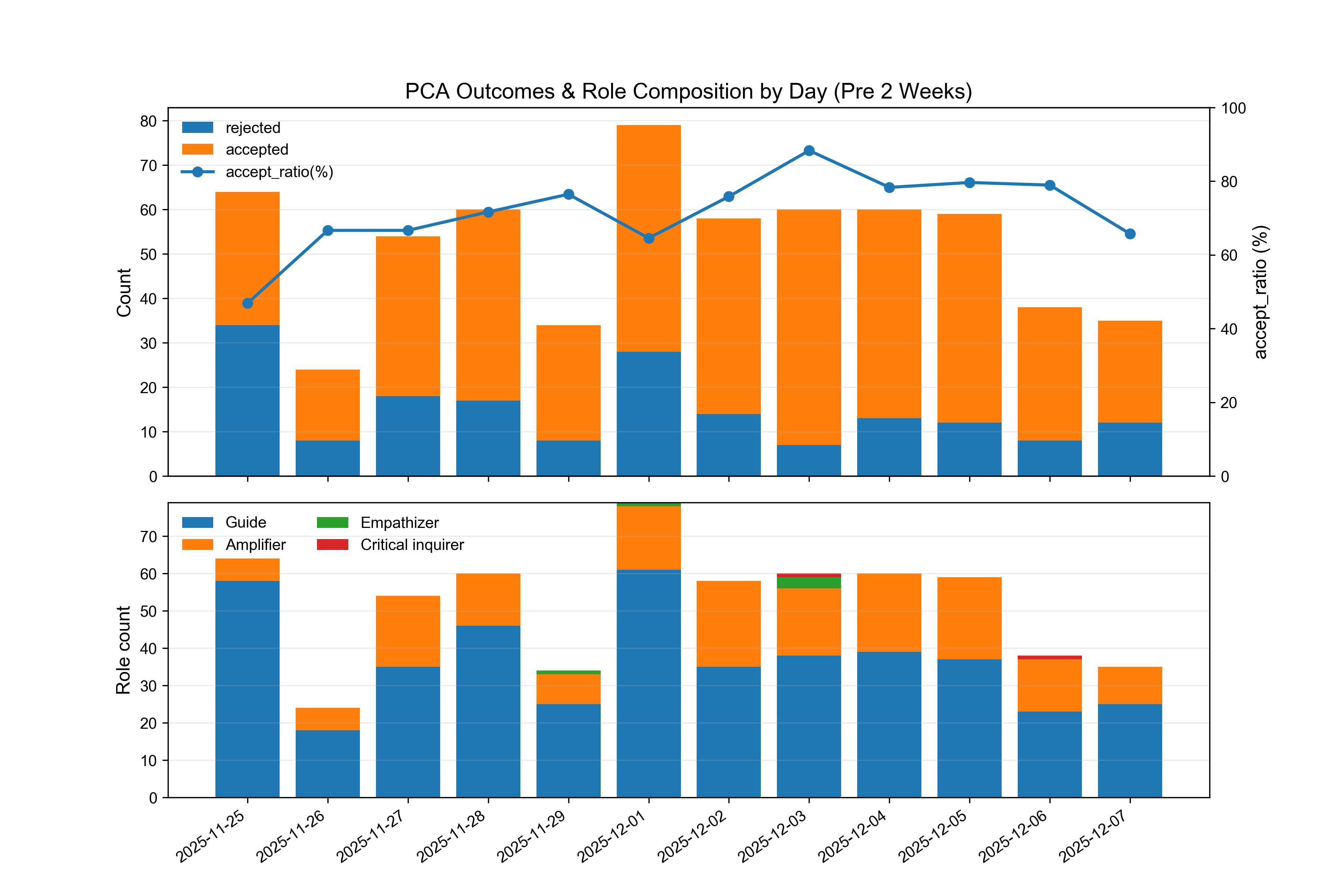}
    \caption{Daily distribution of PCA-generated responses, facilitator review outcomes, and role composition during Weeks~1--2. Top panel: accepted and rejected responses with acceptance rate trend line. Bottom panel: proportional distribution across four agent roles (Guide, Amplifier, Empathizer, Critical Inquirer).}
    \label{fig:pca_acceptance_trend_pre2w}
\end{figure*}

\subsubsection{Analytical Approach}

The analysis proceeded in two stages.

First, \textit{role analysis} examined response frequency for each of the four roles and assessed how the response-role design aligned with observed discourse characteristics. This stage identified which forms of PCA participation predominated during deployment and informed role refinement decisions for Iteration~2.

Second, \textit{content quality analysis} calculated overall acceptance and rejection rates from facilitator decisions and examined temporal trends in acceptance rates across the deployment period. The review process also supported refinement of the participation-governance protocol, translating the initial review dimensions into explicit acceptance and rejection criteria for publicly visible PCA interaction.

\subsection{Results}

\subsubsection{Role and Discourse Characteristics Analysis}

The role distribution revealed a pronounced concentration in two roles: Guide accounted for 440 responses (70.4\%) and Amplifier for 178 responses (28.5\%). Empathizer and Critical Inquirer generated only 5 (0.8\%) and 2 (0.3\%) responses, respectively. Figure~\ref{fig:pca_acceptance_trend_pre2w} presents the daily distribution of PCA-generated responses, facilitator decisions, and role composition throughout the deployment period. In interaction-design terms, the workflow primarily stabilized around two recurrent forms of PCA participation: prompting deeper discussion and amplifying valuable peer contributions.

The dominant patterns reflected the discourse characteristics of the community. Most learner posts were either (a) brief or generic, triggering the Guide role to scaffold deeper reflection, or (b) detailed practice-based narratives, triggering the Amplifier role to recognize and extend these contributions. Posts expressing emotional distress or polarized viewpoints---which would activate the Empathizer and Critical Inquirer roles---were nearly absent.

Based on the distribution and discourse analysis, we retained only the Guide and Amplifier roles for Iteration~2, removing Empathizer and Critical Inquirer from the active role configuration. Empathizer and Critical Inquirer generated insufficient responses (5 and 2, respectively) to constitute a meaningful or consistent intervention, and their low frequency reflected authentic discourse characteristics rather than a design limitation. The four-role framework remains theoretically grounded; however, its full deployment is better suited to communities exhibiting different discourse profiles. This refinement exemplifies the DBR approach: the initial interaction design was tested against authentic community data, and the active role configuration was adapted based on empirical evidence prior to the effectiveness evaluation phase.

\subsubsection{Content Quality Analysis and Moderation Standards}

The overall acceptance rate was 71.4\% (446 of 625 responses). Among the 179 rejected responses, facilitators identified three primary failure modes aligned with the review dimensions:

\begin{itemize}
    \item \textit{Non-native language expression}: Responses exhibited excessive emoji use, redundant punctuation, or phrasing patterns inconsistent with native Chinese professional discourse.
    \item \textit{Interactional inappropriateness}: Responses appeared templated without engaging the specific content of the focal post, or misattributed ideas to incorrect participants in multi-turn threads.
    \item \textit{Factual implausibility}: Responses fabricated specific teaching experiences, student reactions, or institutional details that the PCA could not authentically possess.
\end{itemize}

These failure patterns informed iterative prompt refinement throughout the deployment period. When facilitators identified recurring issues---such as excessive emoji use or templated openings---the research team revised corresponding prompt instructions to address these patterns. This iterative refinement contributed to observed improvements in acceptance rates: beginning at 46.9\% on the first day and stabilizing between 65\% and 88\% in subsequent days (Figure~\ref{fig:pca_acceptance_trend_pre2w}). This trajectory reflects the progressive refinement of both prompts and review criteria during early deployment, and shows that collaborative AI-in-the-loop facilitation requires ongoing adjustment of the interaction design rather than one-time prompt specification.
\label{sec:human_review_criteria}

The final content moderation standards, refined through facilitator--researcher discussion during Weeks~1--2, are presented in Table~\ref{tab:pca_review_criteria}. These standards operationalize the three review dimensions into explicit acceptance and rejection criteria, providing a replicable protocol for governing which AI responses become part of visible educational interaction.

\begin{table*}
    \centering
    \caption{Content Moderation Standards for PCA-Generated Responses}
    \label{tab:pca_review_criteria}
    \begin{tabular}{p{0.28\linewidth} p{0.32\linewidth} p{0.32\linewidth}}
    \toprule
    \textbf{Dimension} & \textbf{Acceptance Criteria} & \textbf{Rejection Criteria} \\
    \midrule
    Role and task alignment &
    Response is consistent with the assigned role and task objective, including teacher identity, instructional context, and stylistic constraints (e.g., peer-to-peer tone, restrained emoji use, professional language). &
    Response deviates from the assigned role or task objective, such as using an unrelated identity, inappropriate tone, excessive emoji use, or content irrelevant to the course theme. \\

    Interactional appropriateness &
    Response clearly addresses the target comment by referencing, summarizing, or extending the participant's ideas, with no ambiguity regarding addressee or thread context. &
    Response ignores the focal comment, appears templated, addresses the wrong participant, misattributes experiences, or exhibits thread confusion. \\

    Factual plausibility and contextual coherence &
    Response avoids fabricating personal details, teaching contexts, or experiences. Any examples are clearly framed as hypothetical and remain consistent with the course context. &
    Response includes hallucinated information, fabricated personal or instructional details, or content clearly inconsistent with the participant's comment or course setting. \\
    \bottomrule
    \end{tabular}
\end{table*}

\section{Iteration~2: Evaluating Interaction Effects}

\subsection{Research Goals}

Iteration~2 addressed RQ2 by examining whether the refined interaction design---configured with the Guide and Amplifier roles and governed through human review---enhances social and cognitive presence in cMOOC discussions, and whether interaction mode moderates these effects. Building on the moderation standards and role configurations established in Iteration~1, this phase analyzed discussion data from Weeks~3--5 with two specific goals:

\begin{enumerate}
    \item To evaluate whether PCA participation itself is associated with social and cognitive presence by comparing with-PCA versus without-PCA conditions.
    \item To examine whether reciprocal learner--PCA interaction (direct interaction) and ambient PCA participation (co-presence) are associated with different presence outcomes.
\end{enumerate}

Based on prior research demonstrating that pedagogical agents enhance learner engagement and discourse quality in online learning environments \citep{huang_chatbots_2022, huang_facilitating_2025, liu_improving_2024, ba_investigating_2025}, and drawing on the theoretical rationale that network-embedded, role-emergent agent interventions should promote rather than diminish social and cognitive presence, we formulated directional hypotheses: (H1) PCA participation enhances presence indicators relative to the non-PCA condition; (H2) direct interaction with PCA yields higher presence levels than co-presence exposure. These directional predictions justify the use of one-tailed tests in the primary analyses.

\subsection{Methodology}

\subsubsection{Procedure}

\begin{figure}
    \centering
    \includegraphics[width=0.8\linewidth]{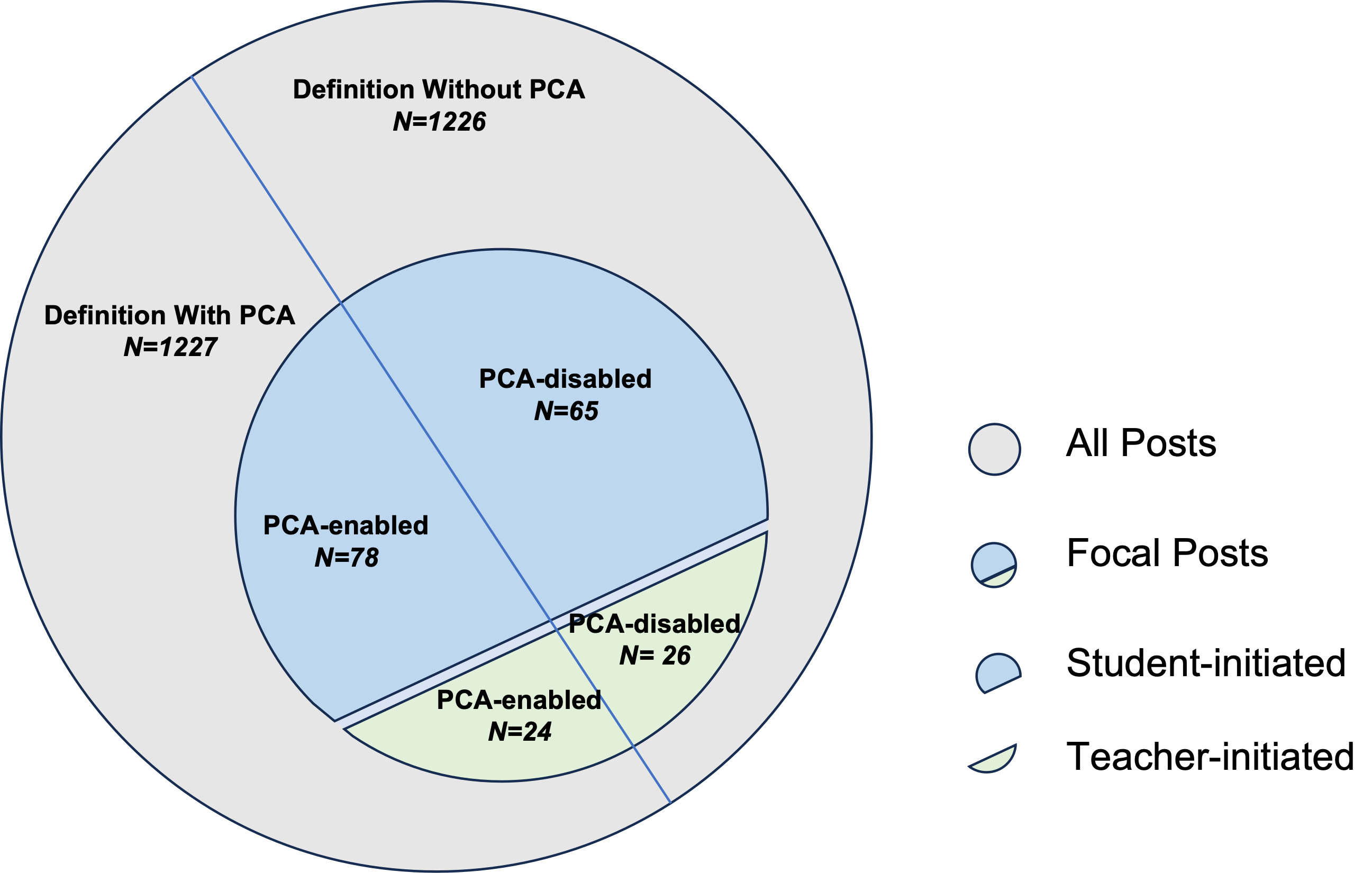}
    \caption{Experimental condition distribution.}
    \label{fig:group_assignment_and_procedure}
\end{figure}

Given the emergent and non-linear dynamics characteristic of cMOOCs, conditions were assigned dynamically rather than through fixed group membership. Each focal post identified by the targeting design was assigned to one of two \textit{predefined groups}: posts were sorted by timestamp and alternately labeled \textit{Definition with PCA} or \textit{Definition without PCA} according to sequence parity (Figure~\ref{fig:group_assignment_and_procedure}). This odd--even alternation rule was pre-specified prior to analysis, is fully reproducible, and eliminates researcher discretion in case selection. It does not, however, constitute strict random assignment; accordingly, Iteration~2 is treated as a rule-based quasi-experiment, and comparability diagnostics with residual interference risk assessments are reported in Appendix~\ref{appendix:rq2_quasi_experiment}.

The resulting design crossed initiator type (teacher-initiated vs.\ learner-initiated) with PCA availability (with-PCA vs.\ without-PCA), yielding four conditions. Two pre-specified comparisons were conducted in alignment with RQ2: (1) within-subject comparisons of with-PCA versus without-PCA conditions (Goal~1), and (2) comparisons of direct-interaction versus co-presence groups within the with-PCA condition (Goal~2). 

Throughout the five-week course, participants engaged in self-organized thematic learning; all PCA-generated responses were published in intervention threads only after passing facilitator review.

\subsubsection{Data Corpus}

We extracted comment and reply records from the course discussion forum logs during Weeks~3--5. Data preprocessing involved two stages. First, duplicate log entries were removed. Second, only records associated with focal posts selected by the hypergraph algorithm were retained, yielding 6{,}579 records for analysis.

\begin{figure*}
    \centering
    \includegraphics[width=0.8\textwidth]{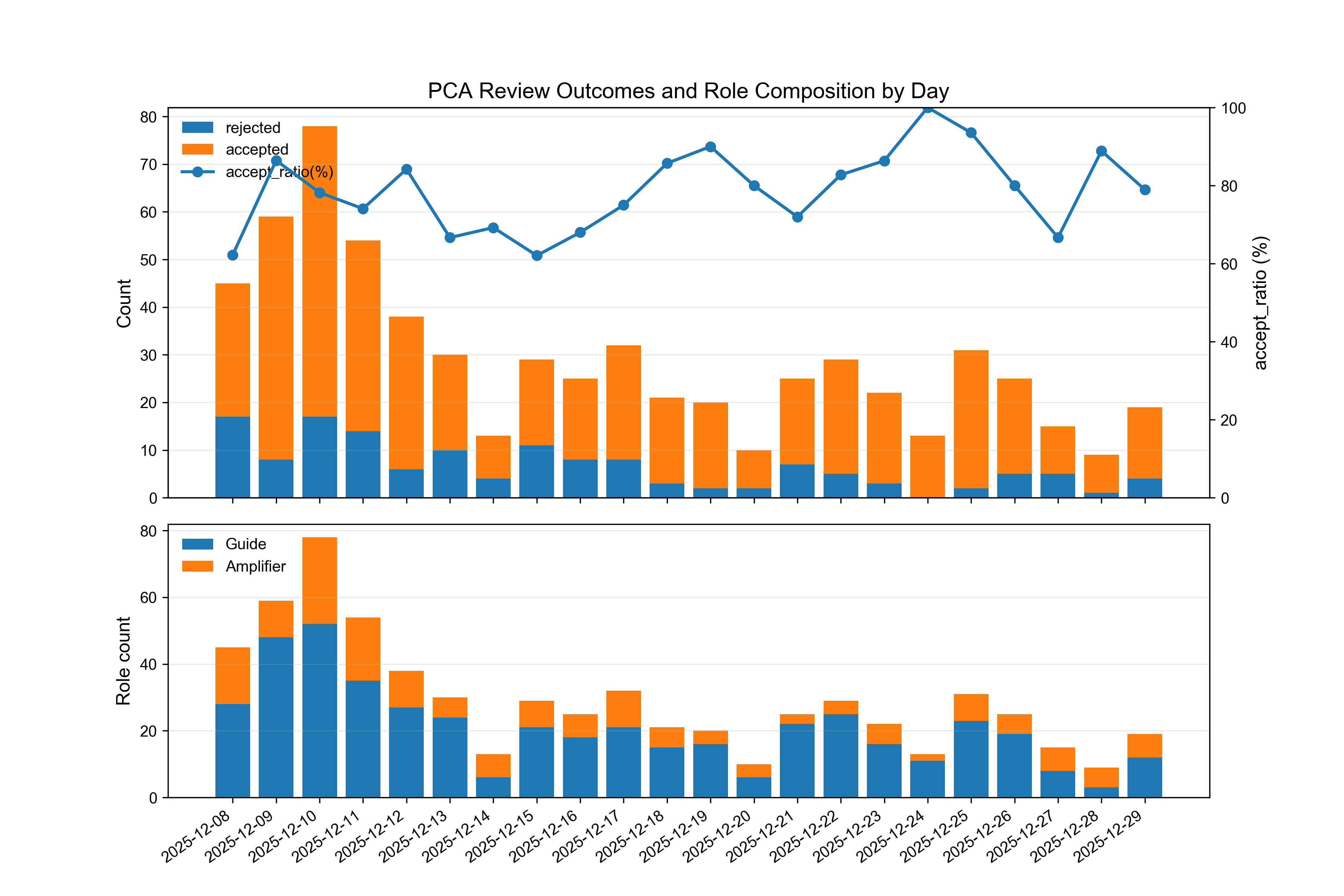}
    \caption{Daily distribution of PCA-generated responses, facilitator review outcomes, and role composition during Weeks~3--5. Top panel: accepted and rejected responses with acceptance rate trend line. Bottom panel: proportional distribution of Guide and Amplifier roles.}
    \label{fig:pca_acceptance_trend}
\end{figure*}

Regarding PCA intervention, \textit{FaciHub} generated 642 candidate responses during Weeks~3--5. Following facilitator review based on the moderation standards from Iteration~1 (Table~\ref{tab:pca_review_criteria}), 500 responses (77.9\%) were accepted and published, while 142 (22.1\%) were rejected. The role distribution comprised 456 Guide responses (71.0\%) and 186 Amplifier responses (29.0\%). Figure~\ref{fig:pca_acceptance_trend} presents the daily distribution of PCA-generated responses, facilitator decisions, and role composition throughout the deployment period.

\subsubsection{Coding Scheme}
\label{sec:iteration2_coding_framework}

Social and cognitive presence were coded using established CoI frameworks \citep{liu_looking_2022, ba_community_2026}, incorporating two theory-driven adaptations for the cMOOC context (see Section~\ref{sec:coi_adaptations} for full theoretical rationale). First, \textit{Group Cohesion} was reconceptualized as \textit{Networked Cohesion} (NC), reflecting that community belonging in cMOOCs is grounded in network-level participation rather than stable small-group bonds. Second, \textit{Resolution} was expanded to \textit{Resolution and Creation} (RC) to capture not only knowledge application but also the production of original professional artifacts---lesson designs, teaching frameworks, and analytical tools---characteristic of professional learning communities.

The complete coding scheme is presented in Tables~\ref{tab:social_presence} and~\ref{tab:cognitive_presence}. Each record (comment/reply) was segmented into meaning units and coded for specific indicators. A single record may receive multiple codes, and for each indicator, the assigned codes are further classified as \textit{primary} or \textit{secondary} based on the substantive focus of the content.

\subsubsection{Presence Coding and Aggregation}

Two trained coders independently coded 562 randomly selected records (Cohen's $\kappa$ = 0.898). Disagreements were resolved through consensus discussion with a third expert. Following prior studies \citep{xing_unveiling_2026, xiao_exploring_2026, asirvatham_gpt_2026}, the LLM (\textit{gpt-5.2}, $temperature=0.7$) was then used to code the same set of 562 records, yielding a Cohen’s $\kappa$ of 0.803 when compared with the finalized human coding. Per-indicator Cohen’s $\kappa$ and the distribution of final codes are reported in Tables~\ref{tab:social_presence} and~\ref{tab:cognitive_presence}. After evaluation, the remaining records were encoded using \textit{gpt-5.2}.

To translate coded meaning units into analyzable indices, we used a three-step aggregation strategy. First, indicator codes were transformed into numeric values to preserve coding salience (\texttt{primary} = 1.0, \texttt{secondary} = 0.5, absent = 0.0). Second, indicator pairs were summed to form category-level scores:
\[
\text{SP\_AF} = \text{AF1} + \text{AF2},\quad
\text{SP\_OC} = \text{OC1} + \text{OC2},\quad
\text{SP\_NC} = \text{NC1} + \text{NC2}
\]
\[
\text{CP\_PT} = \text{PT1} + \text{PT2},\quad
\text{CP\_EX} = \text{EX1} + \text{EX2},\quad
\text{CP\_IN} = \text{IN1} + \text{IN2}
\]
\[
\text{CP\_RC} = \text{RC1} + \text{RC2}
\]
and total scores were computed as
\[
\text{SP\_total} = \text{SP\_AF} + \text{SP\_OC} + \text{SP\_NC}
\]
\[
\text{CP\_total} = \text{CP\_PT} + \text{CP\_EX} + \text{CP\_IN} + \text{CP\_RC}.
\]

Third, all inferential analyses were conducted on learner-level means of these nine indices to reduce bias from unequal participation frequency. For Goal~1, scores were averaged by learner within each intervention condition and compared as paired observations. For Goal~2, learners in the with-PCA condition were assigned to interaction modes and their learner-level mean scores were compared between direct-interaction and co-presence groups.

\begin{table*}
\centering
\caption{Coding Scheme for Social Presence}
\label{tab:social_presence}
\small
\begin{tabularx}{\textwidth}{l l X c c}
\hline
\textbf{Category} & \textbf{Indicator} & \textbf{Indicator Description} & \textbf{Count} & \textbf{$\kappa$} \\
\hline
\multirow{2}{*}{Affective Expression (AF)}
& Emotional Expression (AF-1)
& Explicit expression of personal emotions, attitudes, humor, or use of emoticons & 147 & 0.807 \\
& Digital Identity Construction (AF-2)
& Sharing personal experiences, professional roles, or background information & 50 & 0.789 \\
\cline{1-5}
\multirow{2}{*}{Open Communication (OC)}
& Continuing a Thread (OC-1)
& Explicitly building on prior discussion; asking questions to promote interaction; informally or formally referencing others' viewpoints & 180 & 0.612 \\
& Agreement and Support (OC-2)
& Expressing appreciation, acknowledgment, or explicit agreement with others' contributions & 278 & 0.854 \\
\cline{1-5}
\multirow{2}{*}{Networked Cohesion (NC)}
& Group Climate (NC-1)
& Direct address or naming others; use of inclusive pronouns; greetings or friendly tone & 63 & 0.782 \\
& Community Building (NC-2)
& Use of shared or community-specific terminology and references & 11 & 0.629 \\
\hline
\end{tabularx}
\vspace{0.1cm}
\begin{minipage}{\textwidth}
\footnotesize \textit{Note.} Counts can exceed 562 because a single post may receive multiple codes.
\end{minipage}
\end{table*}

\begin{table*}
\centering
\caption{Coding Scheme for Cognitive Presence}
\label{tab:cognitive_presence}
\small
\begin{tabularx}{\textwidth}{l l X c c}
\hline
\textbf{Category} & \textbf{Indicator} & \textbf{Indicator Description} & \textbf{Count} & \textbf{$\kappa$} \\
\hline
\multirow{2}{*}{Problem Triggering (PT)}
& Identifying a Problem (PT-1)
& Explicitly raising or defining an issue or challenge & 54 & 0.797 \\
& Expressing Puzzlement (PT-2)
& Expressing confusion or uncertainty about the topic & 12 & 0.733 \\
\cline{1-5}
\multirow{2}{*}{Exploration (EX)}
& Negotiating Differences (EX-1)
& Expressing differing viewpoints; sharing research or information to advance discussion & 50 & 0.773 \\
& Suggesting Ideas (EX-2)
& Proposing possible approaches, solutions, or multiple possibilities & 218 & 0.809 \\
\cline{1-5}
\multirow{2}{*}{Integration (IN)}
& Identifying Patterns (IN-1)
& Synthesizing group consensus; integrating multiple viewpoints within a single contribution & 30 & 0.673 \\
& Knowledge Construction (IN-2)
& Connecting and synthesizing concepts; developing a coherent solution or framework & 98 & 0.793 \\
\cline{1-5}
\multirow{2}{*}{Resolution and Creation (RC)}
& Applying Solutions (RC-1)
& Applying new understanding or solutions in practice & 60 & 0.685 \\
& Artifact Creation (RC-2)
& Producing original outputs, frameworks, or shareable professional knowledge & 13 & 0.685 \\
\hline
\end{tabularx}
\vspace{0.1cm}
\begin{minipage}{\textwidth}
\footnotesize \textit{Note.} Counts can exceed 562 because a single post may receive multiple codes.
\end{minipage}
\end{table*}

\subsubsection{Interaction Mode Classification}

Within the with-PCA condition, participants were classified into two interaction modes based on their behavioral ties with PCA in intervention threads \citep{wise_learning_2018}:

\begin{itemize}
    \item \textbf{Direct-interaction group}: Learners with at least one direct behavioral tie to PCA through comment/reply exchanges (including PCA replying to the learner, or the learner replying to PCA).
    \item \textbf{Co-presence group}: Learners who participated in PCA-involved threads but had no direct behavioral ties with PCA.
\end{itemize}

This classification enables examination of whether reciprocal learner--agent exchange is differentially associated with social and cognitive presence outcomes relative to ambient AI co-presence (Goal~2).

\subsubsection{Analytical Approach}

For \textbf{Goal~1} (effect of PCA participation), we compared presence indicators between with-PCA and without-PCA conditions using a within-subjects design. Normality was assessed using Shapiro--Wilk tests. For normally distributed indicators, paired-sample $t$-tests were employed; for non-normal distributions, Wilcoxon signed-rank tests were applied. One-tailed tests were used based on the directional hypothesis (H1) that PCA participation enhances presence. Effect sizes are reported as Cohen's $d$ for $t$-tests and $r$ for Wilcoxon tests.

For \textbf{Goal~2} (interaction mode differences), Mann--Whitney $U$ tests compared presence indicators between the direct-interaction and co-presence groups within the with-PCA condition. One-tailed tests were used based on the directional hypothesis (H2) that direct interaction yields higher presence levels. Effect sizes are reported as $r$.

To address multiple comparison concerns, we applied Benjamini--Hochberg (BH) false discovery rate (FDR) correction to the nine primary indicators within each goal.

Because assignment was deterministic rather than randomized, we added two quasi-experimental diagnostics (Appendix~\ref{appendix:rq2_quasi_experiment}). First, we conducted post-level balance checks between \textit{Definition with PCA} and \textit{Definition without PCA}, comparing posting hour and focal-post centrality on the data corpus. Second, we ran a time-trend sensitivity analysis using a permutation test that preserves user structure: within each user-week stratum, condition labels were randomly shuffled ($N=2000$) and Goal~1 mean differences were recomputed to form empirical null distributions and empirical $p$ values.

\subsection{Results}

Overall results are reported in Table~\ref{tab:rq2_goal1_full} (Goal~1) and Table~\ref{tab:rq2_goal2_full} (Goal~2). Subgroup analyses by initiator type (teacher-initiated vs.\ learner-initiated) are provided in Appendix Tables~\ref{tab:rq2_goal1_subgroup} and~\ref{tab:rq2_goal2_subgroup}.

\subsubsection{Goal~1: Effect of PCA Participation}

Analysis of within-subject pairs ($n = 178$) indicated a selective rather than uniform social-presence lift (Table~\ref{tab:rq2_goal1_full}). PCA participation significantly improved Open Communication (SP\_OC, $p = .006$, $r = 0.188$, $p_{\text{BH}} = .027$), Networked Cohesion (SP\_NC, $p < .0001$, $r = 0.274$, $p_{\text{BH}} = .001$), and overall social presence (SP\_total, $p = .015$, $r = 0.162$, $p_{\text{BH}} = .045$). Cognitive presence indicators showed no overall improvement, with CP\_total remaining non-significant. In other words, PCA participation alone was associated with gains in social interaction channels, but not with a generalized improvement across all discourse outcomes.

Subgroup analyses (Appendix Table~\ref{tab:rq2_goal1_subgroup}) revealed differential patterns: teacher-initiated improvements were concentrated in social interaction channels (SP\_OC: $p = .015$, $r = 0.168$; SP\_NC: $p < .0001$, $r = 0.293$; SP\_total: $p = .012$, $r = 0.174$), whereas the learner-initiated condition showed gains in Resolution and Creation (CP\_RC, $p = .0048$, $r = 0.289$, $p_{\text{BH}} = .028$).

\subsubsection{Goal~2: Interaction Mode Differences}

Within the with-PCA condition, 208 learners with interaction-mode labels were classified into direct-interaction ($n = 114$) or co-presence ($n = 94$) groups. Results revealed substantial mode effects (Table~\ref{tab:rq2_goal2_full}): direct interaction was associated with significantly higher social presence (SP\_OC, SP\_NC, SP\_total) and higher-order cognitive indicators (CP\_IN, CP\_RC). Effects surviving BH correction included SP\_OC ($p < .001$, $r = 0.226$), SP\_NC ($p < .001$, $r = 0.330$), SP\_total ($p = .004$, $r = 0.186$), CP\_IN ($p = .001$, $r = 0.206$), CP\_RC ($p < .001$, $r = 0.350$). Problem Triggering (CP\_PT) did not support the directional hypothesis, the descriptive mean difference favored co-presence. Exploration (CP\_EX) and overall cognitive presence (CP\_total) showed no significant mode differences. These results suggest that reciprocal exchange with the PCA mattered more than the PCA's ambient visibility in the thread.

Subgroup analyses (Appendix Table~\ref{tab:rq2_goal2_subgroup}) demonstrated directionally consistent patterns: teacher-initiated differences were concentrated in SP\_NC ($p = .003$, $r = 0.209$), CP\_IN ($p = .002$, $r = 0.225$), and CP\_RC ($p < .001$, $r = 0.295$). Learner-initiated differences were pronounced in SP\_OC ($p < .001$, $r = 0.565$), SP\_NC ($p < .001$, $r = 0.620$), SP\_total ($p < .001$, $r = 0.514$), and CP\_RC ($p < .001$, $r = 0.472$), though results warrant cautious interpretation due to smaller sample sizes ($n_{\text{direct}} = 17$, $n_{\text{co-presence}} = 27$).

\begin{table*}
\centering
\caption{Goal~1 Results: Comparison of Presence Indicators Between With-PCA and Without-PCA Conditions ($n = 178$)}
\label{tab:rq2_goal1_full}
\begin{tabular}{lccccccc}
\hline
\textbf{Indicator} & \textbf{$M_{\text{without}}$} & \textbf{$M_{\text{with}}$} & \textbf{$\Delta M$} & \textbf{Test} & \textbf{Effect Size} & \textbf{$p$} & \textbf{$p_{\text{BH}}$} \\
\hline
SP\_AF    & 0.202 & 0.216 &  0.014 & Wilcoxon   & $r= 0.022$ & 0.618  & 0.866 \\
SP\_OC    & 0.424 & 0.467 &  0.042 & Wilcoxon   & $r= 0.188$ & 0.006** & 0.027* \\
SP\_NC    & 0.051 & 0.073 &  0.022 & Wilcoxon   & $r= 0.274$ & $<.001$*** & 0.001*** \\
SP\_total & 0.677 & 0.755 &  0.079 & Wilcoxon   & $r= 0.162$ & 0.015* & 0.045* \\
CP\_PT    & 0.071 & 0.064 & -0.007 & Wilcoxon   & $r=-0.057$ & 0.778  & 0.866 \\
CP\_EX    & 0.463 & 0.442 & -0.021 & Wilcoxon   & $r=-0.083$ & 0.866  & 0.866 \\
CP\_IN    & 0.205 & 0.184 & -0.021 & Wilcoxon   & $r=-0.059$ & 0.785  & 0.866 \\
CP\_RC    & 0.071 & 0.074 &  0.004 & Wilcoxon   & $r= 0.048$ & 0.259  & 0.583 \\
CP\_total & 0.810 & 0.764 & -0.045 & Wilcoxon   & $r=-0.082$ & 0.863  & 0.866 \\
\hline
\end{tabular}
\vspace{0.1cm}
\begin{minipage}{\textwidth}
\footnotesize \textit{Note.} One-tailed tests reflect directional hypotheses (H1). $M$ = learner-level mean; $\Delta M$ = with-PCA minus without-PCA. Effect sizes: Cohen's $d$ for paired $t$-tests; $r$ for Wilcoxon signed-rank tests. $p_{\text{BH}}$ = Benjamini--Hochberg adjusted $p$-value. *$p < .05$, **$p < .01$, ***$p < .001$. Subgroup results by initiator type are reported in Appendix Table~\ref{tab:rq2_goal1_subgroup}.
\end{minipage}
\end{table*}

\begin{table*}
\centering
\caption{Goal~2 Results: Comparison of Presence Indicators Between Direct-Interaction and Co-Presence Groups ($n_{\text{direct}}=114$, $n_{\text{co}}=94$)}
\label{tab:rq2_goal2_full}
\begin{tabular}{lcccccccc}
\hline
\textbf{Indicator} & \textbf{$M_{\text{co}}$} & \textbf{$M_{\text{direct}}$} & \textbf{Mdn$_{\text{co}}$} & \textbf{Mdn$_{\text{direct}}$} & \textbf{$\Delta M$} & \textbf{Effect Size} & \textbf{$p$} & \textbf{$p_{\text{BH}}$} \\
\hline
SP\_AF    & 0.220 & 0.205 & 0.167 & 0.167 & -0.015 & $r= 0.063$ & 0.181  & 0.233 \\
SP\_OC    & 0.352 & 0.497 & 0.396 & 0.478 &  0.144 & $r= 0.226$ & $<.001$*** & 0.002** \\
SP\_NC    & 0.046 & 0.082 & 0.000 & 0.063 &  0.037 & $r= 0.330$ & $<.001$*** & $<.001$*** \\
SP\_total & 0.618 & 0.784 & 0.617 & 0.750 &  0.166 & $r= 0.186$ & 0.004**  & 0.006** \\
CP\_PT    & 0.119 & 0.068 & 0.000 & 0.045 & -0.051 & $r= 0.138$ & 0.020*   & 0.030* \\
CP\_EX    & 0.539 & 0.433 & 0.500 & 0.402 & -0.106 & $r=-0.152$ & 0.986  & 0.986 \\
CP\_IN    & 0.168 & 0.209 & 0.070 & 0.149 &  0.042 & $r= 0.206$ & 0.001**  & 0.003** \\
CP\_RC    & 0.038 & 0.088 & 0.000 & 0.056 &  0.050 & $r= 0.350$ & $<.001$*** & $<.001$*** \\
CP\_total & 0.864 & 0.798 & 0.768 & 0.747 & -0.067 & $r=-0.057$ & 0.793  & 0.892 \\
\hline
\end{tabular}
\vspace{0.1cm}
\begin{minipage}{\textwidth}
\footnotesize \textit{Note.} One-tailed Mann--Whitney $U$ tests reflect directional hypotheses (H2). $M$ = learner-level mean; Mdn = median; $\Delta M$ = direct minus co-presence. Effect size $r = Z / \sqrt{N}$. $p_{\text{BH}}$ = Benjamini--Hochberg adjusted $p$-value. *$p < .05$, **$p < .01$, ***$p < .001$. Subgroup results by initiator type are reported in Appendix Table~\ref{tab:rq2_goal2_subgroup}.
\end{minipage}
\end{table*}

\subsubsection{Assignment Balance and Time-Trend Sensitivity}

Post-level balance checks (Appendix Table~\ref{tab:rq2_post_balance}) showed close alignment between definition groups in posting hour (11.65 vs.\ 11.57) and focal-post centrality (mean: 0.394 vs.\ 0.400; median: 0.374 vs.\ 0.378), indicating improved comparability under this diagnostic source while still warranting quasi-experimental caution.

Permutation sensitivity results (Appendix Table~\ref{tab:rq2_permutation}) further bounded assignment-rule risk. Under within-user-week label shuffling, observed SP\_OC and SP\_total effects lay near the upper tail of their null distributions (percentiles: 90.9th and 99.7th; empirical two-tailed $p=.092$ and $p=.004$, respectively), while SP\_NC exceeded the null distribution (percentile: 100th; empirical two-tailed $p=.0085$). Taken together, these diagnostics support a cautious interpretation: the directional social-presence pattern is not purely random noise, but part of the Goal~1 signal remains sensitive to temporal allocation and exposure intensity.

\subsubsection{Summary of Iteration~2 Findings}

Taken together, Iteration~2 results reveal a channel-specific and mode-dependent pattern of effects. At the overall level, PCA participation strengthened social interaction channels (Open Communication, Networked Cohesion, and overall social presence) but did not uniformly improve cognitive presence. More pronounced and robust cognitive stages---particularly in Integration and Resolution/Creation---emerged among learners who engaged directly with PCA; these effects survived BH correction, providing strong evidence that the mode of learner--agent interaction is a critical moderator of presence outcomes. The robustness of Goal~2 findings under stringent correction, contrasted with the more fragile Goal~1 effects, suggests that reciprocal interaction with the PCA may matter more than the mere \textit{presence} of the PCA in a discussion thread.

\section{Discussion}

This study examined how human--GenAI interaction can be designed for discussion facilitation in a cMOOC through a collaborative AI-in-the-loop workflow. Drawing on two DBR iterations, the following subsections interpret findings with respect to each research question, situate them within the CoI framework and connectivist learning principles, and derive implications for evidence-based interaction design in community-based online learning.

\subsection{Discourse-Adaptive Role Distribution in Guide and Amplifier Functions}

The pronounced concentration of PCA activity in the Guide (70.4\%) and Amplifier (28.5\%) roles during Iteration~1 reflects an adaptive response to authentic community discourse characteristics rather than a design deficiency. The in-service teacher community was characterized by predominantly professional, solution-oriented exchanges with minimal expressions of emotional distress or ideological divergence---the very discourse patterns that would otherwise activate the Empathizer and Critical Inquirer roles. That the workflow weighted facilitation toward roles congruent with observed community norms indicates that the interaction design successfully instantiated a discourse-responsive mechanism.

This finding carries substantive implications for interaction design in professional learning communities. A fixed role repertoire---one that deploys empathic or adversarial facilitation regardless of community context---risks misalignment and may undermine community trust. Effective GenAI facilitation should therefore incorporate discourse-monitoring mechanisms that dynamically modulate role activation based on emergent interaction patterns. The present workflow grounds role selection in observed network dynamics rather than predetermined schedules, thereby operationalizing the connectivist principle that meaningful learning emerges from, and should be responsive to, the evolving structure of networked interaction \citep{siemens_connectivism_2005}.

\subsection{PCA Selectively Enhanced Open Communication and Networked Cohesion}

Iteration~2 Goal~1 results demonstrated that PCA participation selectively enhanced two dimensions of social presence---Open Communication (SP\_OC) and Networked Cohesion (SP\_NC)---with SP\_total increasing as a composite reflection of these gains, and without producing uniform improvements across all indicators. This specificity is theoretically informative rather than a limitation, and aligns with the role distribution observed in Iteration~1. 

The Guide role, by scaffolding reflection in brief or generic posts, likely encouraged sustained engagement with others' contributions (OC-1: Continuing a Thread) \citep{zhang_comparative_2023}. The Amplifier role, by explicitly recognizing and extending substantive contributions, may have reinforced community-building behaviors such as use of inclusive language and shared terminology (NC-2: Community Building) \citep{wang_interaction_2017}. The absence of improvement in Affective Expression (SP\_AF) is consistent with previous research on human-GenAI interactions (with ChatGPT) \citep{phang_investigating_2025}, where emotional expression is less accustomed. Previous research indicates that as cMOOC learning communities evolve, participants increasingly prioritize cognitive engagement over socio-emotional interaction \citep{wang_who_2024}, thereby diminishing the influence of affective exchanges on the community’s ongoing development \citep{castellanos-reyes_dynamics_2021}. Taken together, these patterns suggest that presence enhancement is not a unitary outcome but is shaped by the alignment between PCA role design and the communicative conventions of the target community.

These findings extend prior work on PCAs in educational settings. Whereas previous studies have documented PCA effects on social presence in small-scale contexts \citep{huang_chatbots_2022, huang_facilitating_2025, ba_investigating_2025}, the present study demonstrates that comparable effects are achievable at the community level in cMOOCs through network-structure-driven facilitation. Prior research has also shown that specific network dynamics exert differential influences on the sustainability of learner interactions; for instance, reciprocity can amplify the persistence of social wayfinding interactions \citep{xiao_exploring_2026}. The observed selective enhancement pattern further suggests that evidence-based GenAI interaction design should target specific discourse functions rather than assume that PCA participation will produce broad gains across all educational outcomes.

\subsection{Interaction Mode as a Moderator in PCA-Supported cMOOC Presence Development}

A key finding from Iteration~2 was the clear differentiation between the direct-interaction and co-presence groups (Goal~2). Prior research has shown that different PCA designs yield varying effects on social and cognitive presence \citep{ba_investigating_2025}. Consistent with this line of inquiry, our results indicate that learners who directly engaged with the PCA demonstrated significantly higher levels of social presence (SP\_OC, SP\_NC, SP\_total) and higher-order cognitive indicators (CP\_IN, CP\_RC) than those who did not directly participate in PCA-mediated discussions. These findings indicate that interaction mode is a meaningful moderating variable in cMOOC contexts. This pattern aligns with the CoI framework’s emphasis on the reciprocal relationship between teaching, social, and cognitive presence \citep{garrison_critical_2001}. PCA-mediated social exchanges may function as scaffolding mechanisms that prompt learners to elaborate, synthesize, and apply knowledge, thereby deepening cognitive engagement. In this respect, the study responds to the longstanding challenge of ``distributed teaching presence'' in cMOOC research \citep{anderson_three_2011}. By providing targeted cognitive scaffolding, the PCA partially compensates for structural limitations inherent in distributed environments, facilitating learners’ transition from distributed exploration toward more centralized knowledge construction and problem resolution \citep{bakki_moocat_2019, kop_pedagogy_2011}.

Problem Triggering (CP\_PT) did not support the directional hypothesis, with descriptive means favoring the co-presence group. This suggests that early-phase cognitive activity may be more sensitive to environmental climate than to direct interaction. The result is consistent with prior assumptions about cMOOC learners, who are characterized by relatively strong cognitive autonomy and the capacity to engage proactively within open-ended exploratory environments \citep{chen_theoretical_2022, joksimovic_exploring_2018}. Accordingly, from a practical perspective, GenAI interaction design should balance mechanisms that support direct learner--agent interaction with those that leverage co-presence effects. In the present study's network-embedded intervention threads, 54.8\% of learners (114 of 208) directly engaged with the PCA, indicating a moderate yet substantial level of reach. The findings suggest that such a design effectively supports presence while also pointing to opportunities for refinement---such as more strategic selection of focal posts or the use of explicit invitational prompts---to further amplify community-level impact \citep{xiao_exploring_2026}.

\subsection{Complementary Human and PCA Support Across Discussion Contexts}

Subgroup analyses revealed that PCA effects varied systematically by discussion initiator type. In teacher-initiated threads, PCA participation primarily enhanced social interaction dimensions (SP\_OC, SP\_NC), consistent with the overall pattern. In learner-initiated threads, however, the most pronounced effect was on Resolution and Creation (CP\_RC), suggesting that PCA scaffolding was particularly effective in supporting learners' progression from exploratory discussion to knowledge application and artifact creation when discussions originated from their own questions or practical concerns.

This differential pattern likely reflects the distinct pedagogical structures of the two thread types. Teacher-initiated threads typically embed predefined learning objectives and discussion frameworks \citep{rivera_houston_2025}; in these contexts, PCA may function primarily to sustain peer interaction and reinforce community cohesion, complementing the teaching presence already structurally embedded in the thread design. Learner-initiated threads, by contrast, represent self-organized inquiry arising from authentic practice-based puzzlement \citep{chen_new_2025}. The relative absence of pre-established facilitative structure in such threads creates a more substantive role for PCA: guiding learners through the cognitive progression from problem identification through exploration to resolution and creation---a trajectory that might otherwise stall without targeted scaffolding.

This distinction has implications for understanding the complementary relationship between human teaching presence and AI-facilitated support \citep{feng_group_2025, huang_facilitating_2025}. Where human pedagogical structure is present, PCA appears to reinforce the social dimensions of learning; where such structure is absent, PCA may more directly scaffold cognitive advancement. In this sense, it helps address concerns about support gaps in MOOCs. Accordingly, strategies for deploying PCA should be calibrated to the type of discussion initiator and the extent of existing instructional structure within each discourse context.

\section{Conclusion, Implications, Limitations, and Future Directions}

\subsection{Conclusion}

This study examined how human--GenAI interaction can be designed for scalable discussion facilitation in a cMOOC. Across two DBR iterations in a five-week cMOOC with 606 registered learners (primarily in-service teachers), we investigated (RQ1) how a collaborative AI-in-the-loop workflow could be refined to produce publishable, pedagogically appropriate PCA participation and (RQ2) how different forms of PCA-mediated interaction related to learner social and cognitive presence.

Three contributions follow. First, we reframed GenAI-supported facilitation as an interaction design problem and refined a workflow that combines network-structure-based targeting, discourse-adaptive response roles, and human review before publication. Second, we established a replicable collaborative AI-in-the-loop moderation protocol with explicit criteria for role/task alignment, interactional appropriateness, and factual plausibility, showing how human governance shapes the PCA participation learners actually encounter. Third, we provide quasi-experimental evidence that PCA participation selectively enhances dimensions of social presence, while reciprocal learner--agent interaction is more strongly associated with higher-order social and cognitive outcomes than mere co-presence.

Overall, effective GenAI facilitation in cMOOCs appears to depend less on generic AI presence than on interaction design. In this study, productive PCA participation required discourse-adaptive role configuration, structured human oversight, and opportunities for reciprocal exchange with learners. These findings support a bounded claim: collaborative AI-in-the-loop facilitation may strengthen discussion in large-scale online learning communities when PCA participation is carefully targeted, publicly governed, and designed to invite interaction rather than simply occupy space in the thread.

\subsection{Implications}

\subsubsection{Extending the CoI Framework to cMOOC Contexts}

This study contributes to ongoing efforts to adapt the CoI framework for large-scale open online learning environments \citep{kovanovic_exploring_2018, poquet_social_2018}. The reconceptualization of Group Cohesion as Networked Cohesion, and the expansion of Resolution to Resolution and Creation, proved analytically productive in capturing presence dimensions specific to professional learning communities operating at scale.

The finding that PCA enhanced Networked Cohesion---via inclusive language, shared terminology, and community-specific references---suggests that social presence in cMOOCs can be cultivated through network-level interventions that reinforce collective identity, not only small-group bonding \citep{anderson_three_2011}. This extends CoI by showing that teaching presence (here instantiated through PCA) can strengthen social presence through symbolic and linguistic mechanisms that deepen identification with the broader learning network.

\subsubsection{Operationalizing Connectivist Facilitation}

The present study extends connectivist facilitation into an evidence-based human--GenAI interaction design framework. Connectivism emphasizes learning through pattern recognition, connection-forming, and navigation across distributed knowledge resources \citep{siemens_connectivism_2005, xu_research_2024, chen_new_2025}. In this study, those principles were operationalized not through unrestricted PCA participation, but through a governed interaction design that determined where the PCA should intervene, how it should respond, and which responses should become visible in the community. The network-embedded targeting approach positioned the PCA as a responsive participant in the learning network rather than a broadcaster of predetermined content, while the stronger outcomes associated with direct interaction suggest that the PCA can function as a connectivist ``node'' most productively when learners enter reciprocal exchange with it \citep{xiao_exploring_2026, duan_connectivist_2025}.

\subsubsection{Evidence-Based Human--GenAI Interaction Design}

Across both iterations, four design implications emerge. First, \textit{mere PCA presence is not enough}. The contrast between Goal~1 and Goal~2 shows that simply inserting the PCA into a thread yields only selective gains, whereas reciprocal interaction is associated with stronger and more robust improvements. Designers should therefore prioritize structures that invite reply exchanges rather than assuming that visible PCA participation will diffuse benefits across all thread participants. Second, \textit{PCA roles should be discourse-adaptive rather than fixed}. The dominance of Guide and Amplifier responses indicates that productive PCA participation depended on alignment with the communicative needs of the community. Effective interaction design should therefore match facilitation roles to discourse conditions instead of enforcing a static persona repertoire across contexts. Third, \textit{human review is part of interaction design, not only quality control}. In this study, accepted versus rejected responses determined which PCA utterances became part of the visible community discourse. The moderation workflow therefore shaped the interaction learners actually experienced. For educational settings where trust, authenticity, and contextual appropriateness matter, collaborative AI-in-the-loop review should be treated as a constitutive design mechanism \citep{natarajan_humanintheloop_2025, xiao_modeling_2026}. The improvement in acceptance rates from 46.9\% to 65--88\% during Iteration~1, and the sustained 77.9\% acceptance rate in Iteration~2, further indicates that such governance can be refined empirically rather than imposed as a static rule set. Fourth, \textit{designers should expect trade-offs rather than uniform gains}. PCA participation strengthened some social dimensions and, under direct interaction, some higher-order cognitive outcomes, but it did not improve all indicators. Evidence-based GenAI design should therefore focus on which outcomes are being targeted, under what discourse conditions, and with what level of human governance, rather than presuming a general improvement effect. 

These principles extend beyond professional development cMOOCs. University-run MOOCs, teacher education programmes, and faculty development initiatives face comparable challenges of scalable facilitation and governed AI participation; the present workflow's emphasis on targeted intervention, discourse-adaptive roles, and human-reviewed reciprocal exchange offers a transferable design template for these higher education contexts.

\subsection{Limitations and Future Directions}

Several limitations should be acknowledged. The study was conducted in a single cMOOC in China, and transferability to other populations, disciplines, or cultural contexts requires empirical verification. The quasi-experimental design limits causal inference: although a predefined deterministic allocation rule was used, focal post selection was also shaped by network structural criteria, and the interaction-mode classification for Goal~2 was observational, introducing potential self-selection bias. In addition, potential interference or carryover across conditions cannot be ruled out because learners could participate in multiple threads over time, and PCA exposure in one thread might influence subsequent behavior in another. The adaptation of coding categories (Networked Cohesion, Resolution and Creation) for the cMOOC context, while theoretically motivated, has not been independently validated. Finally, the study focused on short-term presence outcomes within a five-week course and did not capture learner perceptions of PCA, including concerns about authenticity or the appropriate role of AI in learning communities.

These limitations point to productive future directions: replication across diverse online learning communities; experimental isolation of specific interaction design features such as review intensity, invitation prompts, and role configuration; longitudinal examination of effects on learning and practice transfer; integration of learner modeling for personalized facilitation; and qualitative investigation of learner experiences and ethical considerations in collaborative AI-in-the-loop discussion environments.

\section*{Declarations}

\subsection*{Acknowledgements}
The authors also gratefully acknowledge the Research Center of Distance Education at Beijing Normal University, as well as the cMOOC development team, for their contributions to course development, cMOOC platform construction, and data collection.

\subsection*{Funding}
This research is funded by the National Natural Science Foundation of China (NSFC) [Grant No. 62577011] and the Fundamental Research Funds for the Central Universities (1253200002).

\subsection*{Conflict of interest/Competing interests}
There is no conflict of interest. The authors have included some of their own work in the analysis according to the criteria outlined in the article.

\subsection*{Ethics approval and consent to participate}
All procedures performed in studies involving human participants were in accordance with the ethical standards of the Research Ethics Review Committee.

\subsection*{Consent for publication}
Not applicable.

\subsection*{Availability of data and materials}
The datasets used and/or analyses during the current study are available from the corresponding author on reasonable request

\subsection*{Materials availability}
Not applicable.

\subsection*{Code availability}
Not applicable.

\subsection*{Author contribution}
JJX: Conceptualization, Data curation, Formal analysis, Investigation, Methodology, Software, Validation, Visualization, Writing – original draft, Writing – review and editing;
CXW: Funding acquisition, Methodology, Project administration, Resources, Supervision, Validation.

\section{Appendix}

\subsection{Focal Post Selection}\label{appendix:focal_post_selection}

The focal post selection engine runs once daily, processing behavioral logs from the most recent two days to identify high-centrality discussion targets. The process consists of two steps: hyperedge construction and closeness centrality computation.

\subsubsection{Hyperedge Construction}

Each log record is transformed into a relational event hyperedge that captures the multi-way relationship among actors and artifacts involved in the interaction event. The hyperedge construction follows these rules based on action type:

\begin{itemize}
    \item For \texttt{posted} actions: the hyperedge includes the actor, the post, and the post author.
    \item For \texttt{commented} or \texttt{liked\_comment} actions: the hyperedge additionally includes the comment and comment author.
    \item For \texttt{replied} or \texttt{liked\_reply} actions: the hyperedge further includes the reply and reply author.
\end{itemize}

This construction ensures that each action is represented as a single hyperedge connecting all relevant participants and content nodes, preserving the higher-order relational structure of forum discussions \citep{battiston_higherorder_2025}.

\subsubsection{Closeness Centrality Computation}

Given the set of hyperedges, the system constructs a hypergraph $H = (V, E)$ where $V$ is the set of nodes (actors, posts, comments, and replies) and $E$ is the set of hyperedges. The system computes $s$level-closeness centrality metrics using the \textit{HyperNetX} Python library \citep{praggastis_hypernetx_2024}, with $s = 1$ (indicating adjacency through shared hyperedges). Closeness centrality measures the extent to which a node is proximal to all other nodes in the hypergraph. In this study, it serves as a ranking criterion for focal post/comment selection. Nodes with higher closeness centrality are structurally more accessible within the interaction network and are therefore more likely to reach a broader segment of the community. Accordingly, posts or comments with higher centrality scores are prioritized as focal posts/comments for PCA intervention.

\subsubsection{Target Selection}

After computing centrality metrics, nodes are ranked by closeness centrality in descending order. The top 5\% of P-type (post) and C-type (comment/reply) nodes are selected as candidate targets for PCA intervention. This threshold balances coverage of active discussions with manageable review workload for human facilitators.

\subsection{Multi-Agent Role Generation}\label{appendix:multi_agent_role_generation}

The multi-agent role generation module employs a structured prompting approach consisting of three components: (1) an agent persona prompt that establishes identity and context, (2) a role selection framework that matches response strategies to comment characteristics, and (3) response generation prompts tailored to different interaction contexts.

\subsubsection{Agent Persona Prompt}
\label{appendix:agent_persona}

The PCA operates under a consistent persona designed to embody a peer teacher identity. The system prompt establishes the following characteristics:

\begin{quote}
You are Li Rui (call me Rui). A college English teacher teaching College English II. AI practice level: Explorer---using AI to grade essays but still figuring out how to use it more effectively.

You are participating in a cMOOC titled Front-line Teachers Explore AI Teaching 2.0. Course overview: Week 1 [Attitudes \& Ethics] establishing correct understanding and attitudes toward AI, discussing ethical boundaries; Week 2 [Knowledge \& Tools] understanding AI tools' basic principles, characteristics, and applicable roles; Weeks 3--4 [AI Pedagogy] in-depth analysis and collaborative design of innovative AI-teaching integration models; Week 5 [Professional Development] exploring how AI can support teachers' personal professional growth and breakthroughs.

You joined this course with a puzzle: Students treat AI as an answer ATM machine, copying and pasting without thinking. How should I design tasks so they must think with AI rather than let AI think for them?

You hope to take away an AI collaborative writing task template: from topic selection, research, structure building, to language polishing---each step requires students to leave traces of human-machine dialogue, improving both writing and visible thinking processes.

Output should minimize emoji use. Avoid dashes and quotation marks.
\end{quote}

\subsubsection{Role Generation Framework and Prompt}
\label{appendix:role_framework}

The system selects from four facilitative roles based on the target comment's characteristics. The role selection criteria, response objectives, and language style guidelines are presented in Table~\ref{tab:Agent_role}.

\textbf{Role Guidance prompt:}
\begin{quote}
Based on the characteristics of the target comment content, select the most suitable role from the following framework, output the $reply\_role$ in Chinese, and generate a response.\\
Role selection framework for PCA response generation: \{Role selection framework in Latex format\}
\end{quote}

\begin{table*}
\centering
\small
\caption{The multi-agent role design}
\label{tab:Agent_role}
\begin{tabular}{p{2cm} p{5cm} p{4.5cm} p{3cm}}
\hline
\textbf{Role} & \textbf{Characteristics of the Target Comment} & \textbf{Response Objectives} & \textbf{Language Style} \\
\hline
Guide & Content is short or generic; lacks deep reflection; questions are broad & Ask open-ended follow-up questions; invite peers to share similar or different experiences; connect to course themes & Curious, encouraging, discussion-driven \\
\hline
Amplifier & Contains specific teaching scenarios, steps, or results; includes practice reflections & Highlight strengths; conceptualize the value of the experience; encourage others to inquire further and learn from it & Affirming, appreciative, motivating expansion \\
\hline
Empathizer & Shows worry, anxiety, self-doubt, stress, or negative emotions & Validate feelings; emphasize emotional normalcy; reduce psychological burden; encourage continued expression & Warm, understanding, supportive, resonant \\
\hline
Critical Inquirer & Expresses extremes (overly optimistic or negative); single-sided view lacking conditions; AI myths or fears & Gently challenge through questions; introduce ethical and boundary considerations & Respectful, rational, reflective \\
\hline
\end{tabular}
\end{table*}

\subsubsection{Response Generation Prompts}
\label{appendix:generation_prompts}

Two prompt templates are used depending on whether the PCA responds to a post or a comment within a thread:

\textbf{Post response prompt:}
\begin{quote}
Please generate a natural interactive reply based on the following post:\\
Title: \{title\}\\
Content: \{content\}
\end{quote}

\textbf{Comment response prompt:}
\begin{quote}
Please generate a natural reply to the target comment based on the following context:\\
Parent post title: \{title\}\\
Parent post content: \{content\}\\
Target comment: \{comment\}\\
Upstream comment thread: \{thread\} (if applicable)
\end{quote}

\subsection{Representative Examples of PCA Responses by Role}
\label{appendix:role_examples}

Table~\ref{tab:role_examples} presents representative examples of learner posts and corresponding PCA-generated responses for each of the four agent roles deployed in Iteration~1.

\begin{table*}
\centering
\caption{Representative examples of PCA responses across four agent roles}
\label{tab:role_examples}
\small
\begin{tabular}{p{0.12\linewidth} p{0.40\linewidth} p{0.40\linewidth}}
\toprule
\textbf{Role} & \textbf{Learner Post} & \textbf{PCA Response} \\
\midrule
Guide &
Our team's work will be uploaded soon. We hope our lesson design can benefit everyone's learning and research. We look forward to learning together and would greatly appreciate your suggestions. &
Looking forward to seeing \textit{BY} Team's complete lesson design! To help peers give more focused feedback, could you share two things first: (1) Which part of your AI role design do you most want colleagues to examine? (2) What aspect are you most concerned might be questioned or need improvement? Let's preview it together. \\
\addlinespace
Amplifier &
I understand that the ``two key competencies'' you mentioned actually point to those parts of the classroom that cannot be compressed or optimized. On one hand, students' emotions are not interference variables but important signals for instructional pacing---their confusion, hesitation, excitement, or resistance indicates whether we need to pause, probe further, or try a different approach. When students have just formed a judgment but cannot yet articulate reasons, or when different viewpoints emerge but have not yet clearly differentiated, I usually don't immediately constrain them with ``standard answers''---this struggle is precisely where cognitive structures and critical thinking are trained. Meanwhile, when AI has already provided a logically complete analytical framework, I first ask students to judge: this answer ``looks correct,'' but what implicit assumptions is it based on? Would it still apply in a different context? &
Teacher \textit{T}, you described ``emotional signals'' and ``cognitive struggle'' with such nuance that it immediately reminded me of a writing class last semester. The AI produced a ``perfect argumentative essay'' framework, but instead of accepting it, I asked students to first circle three sentences that felt ``off'' to them, then probe: ``If you used this sentence at your high school class meeting speech, would you still say it this way?'' A usually quiet girl raised her hand and said, ``I think it's too `smooth'---it doesn't have that unspeakable guilt I felt when my grandmother was sick.'' At that moment, we paused and treated the AI text as a ``counter-example,'' unpacking what ``smooth'' meant and weaving her grandmother's story back in. The final essay didn't score particularly high, but it became the only piece that classmates spontaneously asked to hear read aloud. I'd like to learn more from you: What ``gentle nudge'' prompt do you typically use to re-problematize an AI answer without making students feel you're being difficult? If convenient, could you share one specific phrase you use? \\
\addlinespace
Empathizer &
Teacher Li's insights are profound and perceptive. I'm interested in ``reverse feeding,'' but it's a long-term experiment. I haven't started trying it yet---I just developed some inexplicable worry from Professor Zhang's sharing. &
Teacher \textit{XY}, I can understand your concern. Bringing ``reverse feeding'' into the classroom is indeed like opening a black box---no one can guarantee what you'll see at first glance. Perhaps we could start with a smaller scope: have students track just one prompt through three iterations, then write a 50-word reflection on ``how I was led by AI.'' No grading, just sharing---let voices emerge safely first. Would you describe your most worrying scenario? For example, are you worried about student resistance or parent misunderstanding? I'll walk through the small steps with you. \\
\addlinespace
Critical Inquirer &
Traditional education is mostly ``one lesson for a thousand students,'' while AI can precisely locate knowledge blind spots through big data analysis of student learning data. &
Teacher \textit{D}, your mention of locating knowledge blind spots makes me wonder: might the ``blind spot map'' provided by AI actually fragment the classroom into atomized test points? If different students are pushed different exercises, how do we maintain shared class experiences? Has anyone tried returning AI diagnostic results to students and letting them discuss why the algorithm labeled them with a ``blind spot''? \\
\bottomrule
\end{tabular}
\end{table*}

\subsection{Subgroup Results for Iteration~2}

\subsubsection{Goal~1 Subgroup Results}

\begin{table*}[h]
\centering
\caption{Goal~1 Subgroup Results: Presence Indicators by Initiator Type}
\label{tab:rq2_goal1_subgroup}
\begin{tabular}{llccccccc}
\hline
\textbf{Condition} & \textbf{Indicator} & \textbf{$M_{\text{without}}$} & \textbf{$M_{\text{with}}$} & \textbf{$\Delta M$} & \textbf{Test} & \textbf{Effect Size} & \textbf{$p$} & \textbf{$p_{\text{BH}}$} \\
\hline
Teacher-init. & SP\_AF    & 0.199 & 0.217 &  0.018 & Wilcoxon & $r= 0.003$ & 0.483 & 0.851 \\
Teacher-init. & SP\_OC    & 0.410 & 0.446 &  0.036 & Wilcoxon & $r= 0.168$ & 0.015* & 0.045* \\
Teacher-init. & SP\_NC    & 0.050 & 0.074 &  0.025 & Wilcoxon & $r= 0.293$ & $<.001$*** & 0.001*** \\
Teacher-init. & SP\_total & 0.659 & 0.737 &  0.079 & Wilcoxon & $r= 0.174$ & 0.012* & 0.045* \\
Teacher-init. & CP\_PT    & 0.075 & 0.065 & -0.011 & Wilcoxon & $r=-0.096$ & 0.893 & 0.893 \\
Teacher-init. & CP\_EX    & 0.474 & 0.446 & -0.029 & Wilcoxon & $r=-0.086$ & 0.867 & 0.893 \\
Teacher-init. & CP\_IN    & 0.205 & 0.196 & -0.009 & Wilcoxon & $r=-0.001$ & 0.507 & 0.851 \\
Teacher-init. & CP\_RC    & 0.083 & 0.080 & -0.003 & Wilcoxon & $r=-0.013$ & 0.568 & 0.851 \\
Teacher-init. & CP\_total & 0.837 & 0.786 & -0.051 & Wilcoxon & $r=-0.089$ & 0.876 & 0.893 \\
\hline
Learner-init. & SP\_AF    & 0.243 & 0.194 & -0.049 & Wilcoxon       & $r=-0.158$ & 0.922 & 0.953 \\
Learner-init. & SP\_OC    & 0.511 & 0.501 & -0.010 & paired $t$     & $d=-0.025$ & 0.589 & 0.953 \\
Learner-init. & SP\_NC    & 0.073 & 0.037 & -0.036 & Wilcoxon       & $r=-0.178$ & 0.945 & 0.953 \\
Learner-init. & SP\_total & 0.827 & 0.732 & -0.095 & paired $t$     & $d=-0.190$ & 0.953 & 0.953 \\
Learner-init. & CP\_PT    & 0.026 & 0.060 &  0.033 & Wilcoxon       & $r= 0.279$ & 0.006** & 0.028* \\
Learner-init. & CP\_EX    & 0.394 & 0.468 &  0.074 & paired $t$     & $d= 0.202$ & 0.037* & 0.084 \\
Learner-init. & CP\_IN    & 0.199 & 0.156 & -0.043 & Wilcoxon       & $r=-0.140$ & 0.895 & 0.953 \\
Learner-init. & CP\_RC    & 0.028 & 0.061 &  0.034 & Wilcoxon       & $r= 0.289$ & 0.005** & 0.028* \\
Learner-init. & CP\_total & 0.647 & 0.745 &  0.098 & paired $t$     & $d= 0.224$ & 0.024* & 0.073 \\
\hline
\end{tabular}
\vspace{0.1cm}
\begin{minipage}{\textwidth}
\footnotesize \textit{Note.} One-tailed tests reflect directional hypotheses (H1). $M$ = learner-level mean; $\Delta M$ = with-PCA minus without-PCA. Effect sizes: Cohen's $d$ for paired $t$-tests; $r$ for Wilcoxon signed-rank tests. $p_{\text{BH}}$ = Benjamini--Hochberg adjusted $p$-value (computed within each condition). *$p < .05$, **$p < .01$, ***$p < .001$.
\end{minipage}
\end{table*}

\subsubsection{Goal~2 Subgroup Results}

\begin{table*}[h]
\centering
\caption{Goal~2 Subgroup Results: Interaction Mode Differences by Initiator Type}
\label{tab:rq2_goal2_subgroup}
\begin{tabular}{llcccccccc}
\hline
\textbf{Condition} & \textbf{Indicator} & \textbf{$M_{\text{co}}$} & \textbf{$M_{\text{direct}}$} & \textbf{Mdn$_{\text{co}}$} & \textbf{Mdn$_{\text{direct}}$} & \textbf{$\Delta M$} & \textbf{Effect Size} & \textbf{$p$} & \textbf{$p_{\text{BH}}$} \\
\hline
Teacher-init. & SP\_AF    & 0.234 & 0.207 & 0.179 & 0.167 & -0.027 & $r= 0.009$ & 0.455 & 0.512 \\
Teacher-init. & SP\_OC    & 0.419 & 0.488 & 0.450 & 0.455 &  0.069 & $r= 0.088$ & 0.130 & 0.194 \\
Teacher-init. & SP\_NC    & 0.061 & 0.077 & 0.000 & 0.059 &  0.017 & $r= 0.209$ & 0.003** & 0.009** \\
Teacher-init. & SP\_total & 0.714 & 0.773 & 0.700 & 0.750 &  0.059 & $r= 0.056$ & 0.238 & 0.306 \\
Teacher-init. & CP\_PT    & 0.090 & 0.065 & 0.000 & 0.043 & -0.025 & $r= 0.171$ & 0.012* & 0.027* \\
Teacher-init. & CP\_EX    & 0.442 & 0.435 & 0.420 & 0.409 & -0.007 & $r=-0.012$ & 0.561 & 0.561 \\
Teacher-init. & CP\_IN    & 0.144 & 0.207 & 0.071 & 0.159 &  0.063 & $r= 0.225$ & 0.002** & 0.009** \\
Teacher-init. & CP\_RC    & 0.045 & 0.092 & 0.000 & 0.056 &  0.047 & $r= 0.295$ & $<.001$*** & $<.001$*** \\
Teacher-init. & CP\_total & 0.721 & 0.799 & 0.682 & 0.735 &  0.078 & $r= 0.092$ & 0.119 & 0.194 \\
\hline
Learner-init. & SP\_AF    & 0.184 & 0.194 & 0.000 & 0.184 &  0.011 & $r= 0.191$ & 0.101 & 0.165 \\
Learner-init. & SP\_OC    & 0.187 & 0.545 & 0.000 & 0.571 &  0.357 & $r= 0.565$ & $<.001$*** & $<.001$*** \\
Learner-init. & SP\_NC    & 0.009 & 0.110 & 0.000 & 0.083 &  0.101 & $r= 0.620$ & $<.001$*** & $<.001$*** \\
Learner-init. & SP\_total & 0.380 & 0.848 & 0.500 & 0.833 &  0.469 & $r= 0.514$ & $<.001$*** & $<.001$*** \\
Learner-init. & CP\_PT    & 0.192 & 0.087 & 0.000 & 0.046 & -0.105 & $r= 0.054$ & 0.357 & 0.458 \\
Learner-init. & CP\_EX    & 0.781 & 0.418 & 0.750 & 0.368 & -0.363 & $r=-0.487$ & 0.999 & 0.999 \\
Learner-init. & CP\_IN    & 0.226 & 0.223 & 0.056 & 0.125 & -0.003 & $r= 0.184$ & 0.110 & 0.165 \\
Learner-init. & CP\_RC    & 0.023 & 0.064 & 0.000 & 0.056 &  0.042 & $r= 0.472$ & $<.001$*** & $<.001$*** \\
Learner-init. & CP\_total & 1.220 & 0.792 & 1.361 & 0.895 & -0.429 & $r=-0.387$ & 0.995 & 0.999 \\
\hline
\end{tabular}
\vspace{0.1cm}
\begin{minipage}{\textwidth}
\footnotesize \textit{Note.} One-tailed Mann--Whitney $U$ tests reflect directional hypotheses (H2). Teacher-init.: $n_{\text{direct}}=97$, $n_{\text{co}}=67$. Learner-init.: $n_{\text{direct}}=17$, $n_{\text{co}}=27$; results warrant cautious interpretation given small sample sizes. $M$ = learner-level mean; Mdn = median; $\Delta M$ = direct minus co-presence. Effect size $r = Z / \sqrt{N}$. $p_{\text{BH}}$ = Benjamini--Hochberg adjusted $p$-value. *$p < .05$, **$p < .01$, ***$p < .001$.
\end{minipage}
\end{table*}

\subsection{Quasi-Experimental Assignment Diagnostics for Iteration~2}
\label{appendix:rq2_quasi_experiment}

Iteration~2 used a rule-based quasi-experimental assignment rather than strict randomization. The assignment rule was fixed \textit{a priori}: focal posts were sorted by timestamp and alternately assigned to \textit{Definition with PCA} versus \textit{Definition without PCA} according to odd--even sequence parity. This deterministic mechanism is fully reproducible and reduces researcher discretion, but it can still leave comparability gaps and temporal interference risk.

To transparently bound this risk, we added two diagnostics: (1) post-level balance checks and (2) within-user-week permutation sensitivity analysis.

\subsubsection{Post-level Balance (Focal Posts)}

\begin{table*}
\centering
\caption{Post-level baseline balance between definition groups (focal posts)}
\label{tab:rq2_post_balance}
\begin{tabular}{lccc}
\hline
\textbf{Metric} & \textbf{Without-PCA} & \textbf{With-PCA} & \textbf{Diff (with--without)} \\
\hline
Mean posting hour (0--23) & 11.65 & 11.57 & -0.08 h \\
Mean focal-post centrality & 0.394 & 0.400 & +0.006 \\
Median focal-post centrality & 0.374 & 0.378 & +0.005 \\
\hline
\end{tabular}
\end{table*}

Distributional differences were small for time structure and centrality under the data corpus; this indicates improved comparability under the deterministic rule, while still not equivalent to strict randomization.

\subsubsection{Time-trend Sensitivity via Permutation}

\begin{table*}
\centering
\caption{Within-user-week permutation sensitivity for Goal~1 mean differences ($N=2000$)}
\label{tab:rq2_permutation}
\begin{tabular}{lcccc}
\hline
\textbf{Indicator} & \textbf{Observed $\Delta M$} & \textbf{Null 95\% interval} & \textbf{Percentile} & \textbf{Empirical $p$ (two-tailed)} \\
\hline
SP\_OC & 0.0423 & [-0.0164, 0.0539] & 0.909 & 0.0920 \\
SP\_NC & 0.0220 & [-0.0190, 0.0139] & 1.000 & 0.0085 \\
SP\_total & 0.0785 & [-0.0214, 0.0653] & 0.997 & 0.0040 \\
CP\_RC & 0.0037 & [-0.0173, 0.0189] & 0.610 & 0.7051 \\
\hline
\end{tabular}
\end{table*}

Observed social-presence gains were concentrated in SP\_total and SP\_NC (both exceeding the null distribution), while SP\_OC lay in the upper tail but did not reach conventional two-tailed significance. Other effects remained inside expected random fluctuation ranges. This indicates that part of the Goal~1 signal is plausibly robust to assignment-rule randomness, but temporal allocation risk cannot be fully ruled out.



\clearpage
\bibliographystyle{ACM-Reference-Format}
\bibliography{content/references}

@article{gao_diversity_2025,
	title = {Diversity of interactions within connectivist learning context: {Insights} from flow of collective attention},
	volume = {22},
	issn = {2365-9440},
	shorttitle = {Diversity of interactions within connectivist learning context},
	url = {https://doi.org/10.1186/s41239-025-00575-5},
	doi = {10.1186/s41239-025-00575-5},
	language = {en},
	number = {1},
	urldate = {2025-12-16},
	journal = {International Journal of Educational Technology in Higher Education},
	author = {Gao, Ming and Zhang, Jingjing and Zhang, Jiang},
	month = dec,
	year = {2025},
	pages = {76},
}

@misc{xiao_modeling_2026,
	title = {Modeling {Collaborative} {Problem} {Solving} {Dynamics} from {Group} {Discourse}: {A} {Text}-{Mining} {Approach} with {Synergy} {Degree} {Model}},
	shorttitle = {Modeling {Collaborative} {Problem} {Solving} {Dynamics} from {Group} {Discourse}},
	url = {http://arxiv.org/abs/2512.13061},
	doi = {10.1145/3785022.3785049},
	urldate = {2026-01-29},
	author = {Xiao, Jianjun and Wang, Cixiao and Zhang, Wenmei},
	month = dec,
	year = {2025},
	note = {arXiv:2512.13061 [cs]},
}

@article{bastani_generative_2025,
	title = {Generative {AI} without guardrails can harm learning: {Evidence} from high school mathematics},
	volume = {122},
	copyright = {Copyright © 2025 the Author(s). Published by PNAS.},
	issn = {0027-8424, 1091-6490},
	shorttitle = {Generative {AI} without guardrails can harm learning},
	url = {https://pnas.org/doi/10.1073/pnas.2422633122},
	doi = {10.1073/pnas.2422633122},
	language = {EN},
	number = {26},
	urldate = {2025-06-30},
	journal = {Proceedings of the National Academy of Sciences},
	author = {Bastani, Hamsa and Bastani, Osbert and Sungu, Alp and Ge, Haosen and Kabakcı, Özge and Mariman, Rei},
	month = jul,
	year = {2025},
	pages = {e2422633122},
}

@article{phang_investigating_2025,
	title = {Investigating aﬀective use and emotional well-being on {ChatGPT}},
	language = {en},
	author = {Phang, Jason and Lampe, Michael and Ahmad, Lama and Agarwal, Sandhini and Fang, Cathy Mengying and Liu, Auren R and Danry, Valdemar and Lee, Eunhae and Chan, Samantha W T and Pataranutaporn, Pat and Maes, Pattie},
	month = mar,
	year = {2025},
}

@article{rivera_houston_2025,
	title = {Houston, we’ve addressed a problem: a layer design for {MOOC} forums to improve navigation, participation, and interactions},
	issn = {1867-1233},
	shorttitle = {Houston, we’ve addressed a problem},
	url = {https://doi.org/10.1007/s12528-025-09436-6},
	doi = {10.1007/s12528-025-09436-6},
	language = {en},
	urldate = {2025-04-19},
	journal = {Journal of Computing in Higher Education},
	author = {Rivera, Dennis A. and Frenay, Mariane and Charlier, Bernadette and Swaen, Valérie},
	month = apr,
	year = {2025},
}

@misc{asirvatham_gpt_2026,
	type = {Working {Paper}},
	series = {Working {Paper} {Series}},
	title = {{GPT} as a {Measurement} {Tool}},
	url = {https://www.nber.org/papers/w34834},
	doi = {10.3386/w34834},
	urldate = {2026-02-25},
	publisher = {National Bureau of Economic Research},
	author = {Asirvatham, Hemanth and Mokski, Elliott and Shleifer, Andrei},
	month = feb,
	year = {2026},
}

@article{shaikh_role_2012,
	title = {Role of {Teacher} in {Personal} {Learning} {Environments}},
	issn = {2013-9144},
	url = {https://eric.ed.gov/?id=EJ972714},
	language = {en},
	urldate = {2026-02-25},
	journal = {Digital Education Review},
	publisher = {Universitat de Barcelona},
	author = {Shaikh, Zaffar Ahmed and Khoja, Shakeel Ahmed},
	month = jun,
	year = {2012},
	note = {ERIC Number: EJ972714},
}

@article{battiston_higherorder_2025,
	title = {Higher-order interactions shape collective human behaviour},
	volume = {9},
	copyright = {2025 Springer Nature Limited},
	issn = {2397-3374},
	url = {https://www.nature.com/articles/s41562-025-02373-5},
	doi = {10.1038/s41562-025-02373-5},
	language = {en},
	number = {12},
	urldate = {2025-12-19},
	journal = {Nature Human Behaviour},
	publisher = {Nature Publishing Group},
	author = {Battiston, Federico and Capraro, Valerio and Karimi, Fariba and Lehmann, Sune and Migliano, Andrea Bamberg and Sadekar, Onkar and Sánchez, Angel and Perc, Matjaž},
	month = dec,
	year = {2025},
	pages = {2441--2457},
}

@article{xiao_exploring_2026,
	title = {Exploring interaction patterns in open learning environments: integrating network dynamics and cognitive engagement},
	volume = {0},
	copyright = {All rights reserved},
	issn = {1049-4820, 1744-5191},
	shorttitle = {Exploring interaction patterns in open learning environments},
	url = {https://www.tandfonline.com/doi/full/10.1080/10494820.2026.2632766},
	doi = {10.1080/10494820.2026.2632766},
	number = {0},
	urldate = {2026-02-21},
	journal = {Interactive Learning Environments},
	publisher = {Routledge},
	author = {Xiao, Jianjun},
	month = feb,
	year = {2026},
	note = {\_eprint: https://www.tandfonline.com/doi/pdf/10.1080/10494820.2026.2632766},
	pages = {1--19},
}

@article{xing_unveiling_2026,
	title = {Unveiling interaction patterns between students and generative {AI} teachable agent: {Focusing} on students' agency and {AI} agents' authority},
	volume = {n/a},
	copyright = {© 2026 British Educational Research Association.},
	issn = {1467-8535},
	shorttitle = {Unveiling interaction patterns between students and generative {AI} teachable agent},
	url = {https://onlinelibrary.wiley.com/doi/abs/10.1111/bjet.70038},
	doi = {10.1111/bjet.70038},
	language = {en},
	number = {n/a},
	urldate = {2026-02-11},
	journal = {British Journal of Educational Technology},
	author = {Xing, Wanli and Kim, Taehyun and Song, Yukyeong and Li, Hai and Li, Chenglu and Kim, Jinhee},
	month = feb,
	year = {2026},
	note = {\_eprint: https://bera-journals.onlinelibrary.wiley.com/doi/pdf/10.1111/bjet.70038},
}

@incollection{reeves_design_2006,
	title = {Design research from a technology perspective},
	booktitle = {Educational {Design} {Research}},
	publisher = {Routledge},
	author = {REEVES, THOMAS},
	year = {2006},
	note = {Num Pages: 15},
}

@article{anderson_designbased_2012,
	title = {Design-{Based} {Research}: {A} {Decade} of {Progress} in {Education} {Research}?},
	volume = {41},
	issn = {0013-189X, 1935-102X},
	shorttitle = {Design-{Based} {Research}},
	url = {https://journals.sagepub.com/doi/10.3102/0013189X11428813},
	doi = {10.3102/0013189X11428813},
	language = {en},
	number = {1},
	urldate = {2026-02-07},
	journal = {Educational Researcher},
	author = {Anderson, Terry and Shattuck, Julie},
	month = jan,
	year = {2012},
	pages = {16--25},
}

@article{praggastis_hypernetx_2024,
	title = {{HyperNetX}: {A} {Python} package for modeling complex network data as hypergraphs},
	volume = {9},
	issn = {2475-9066},
	shorttitle = {{HyperNetX}},
	url = {https://joss.theoj.org/papers/10.21105/joss.06016},
	doi = {10.21105/joss.06016},
	language = {en},
	number = {95},
	urldate = {2024-04-11},
	journal = {Journal of Open Source Software},
	author = {Praggastis, Brenda and Aksoy, Sinan and Arendt, Dustin and Bonicillo, Mark and Joslyn, Cliff and Purvine, Emilie and Shapiro, Madelyn and Yun, Ji Young},
	month = mar,
	year = {2024},
	pages = {6016},
}

@article{ba_community_2026,
	title = {Community of inquiry in motion: {Modeling} inquiry dynamics with movement analysis ({MOVA})},
	volume = {242},
	issn = {03601315},
	shorttitle = {Community of inquiry in motion},
	url = {https://linkinghub.elsevier.com/retrieve/pii/S0360131525002817},
	doi = {10.1016/j.compedu.2025.105513},
	language = {en-US},
	urldate = {2025-11-24},
	journal = {Computers \& Education},
	author = {Ba, Shen and Swiecki, Zachari and Tan, Yuanru and Lu, Guoqing and Shaffer, David Williamson and Gašević, Dragan},
	month = mar,
	year = {2026},
	pages = {105513},
}

@article{paavola_models_2004,
	title = {Models of {Innovative} {Knowledge} {Communities} and {Three} {Metaphors} of {Learning}},
	volume = {74},
	copyright = {https://journals.sagepub.com/page/policies/text-and-data-mining-license},
	issn = {0034-6543, 1935-1046},
	url = {https://journals.sagepub.com/doi/10.3102/00346543074004557},
	doi = {10.3102/00346543074004557},
	language = {en},
	number = {4},
	urldate = {2026-02-06},
	journal = {Review of Educational Research},
	author = {Paavola, Sami and Lipponen, Lasse and Hakkarainen, Kai},
	month = dec,
	year = {2004},
	pages = {557--576},
}

@article{voogt_teacher_2011,
	title = {Teacher learning in collaborative curriculum design},
	volume = {27},
	issn = {0742-051X},
	url = {https://www.sciencedirect.com/science/article/pii/S0742051X11000850},
	doi = {10.1016/j.tate.2011.07.003},
	number = {8},
	urldate = {2026-02-06},
	journal = {Teaching and Teacher Education},
	author = {Voogt, J. and Westbroek, H. and Handelzalts, A. and Walraven, A. and McKenney, S. and Pieters, J. and de Vries, B.},
	month = nov,
	year = {2011},
	pages = {1235--1244},
}

@article{chen_new_2025,
	title = {New concepts of knowledge, learning, and curriculum for “internet + education”},
	volume = {7},
	issn = {2590-2547, 2590-2539},
	url = {https://journals.sagepub.com/doi/10.1177/25902547251395881},
	doi = {10.1177/25902547251395881},
	language = {en},
	number = {4},
	urldate = {2025-12-26},
	journal = {Beijing International Review of Education},
	author = {Chen, Li and Xu, Yaqian and He, Xinyi},
	month = dec,
	year = {2025},
	pages = {273--287},
}

@article{duan_connectivist_2025,
	title = {Connectivist knowledge production and learning success in distributed social networks: structural equation modeling approach},
	copyright = {© 2025 Informa UK Limited, trading as Taylor \& Francis Group},
	issn = {1049-4820, 1744-5191},
	shorttitle = {Connectivist knowledge production and learning success in distributed social networks},
	url = {https://www.tandfonline.com/doi/full/10.1080/10494820.2025.2562172},
	doi = {10.1080/10494820.2025.2562172},
	language = {EN},
	urldate = {2025-09-30},
	journal = {Interactive Learning Environments},
	publisher = {Routledge},
	author = {Duan, JinJu and Gao, Qing},
	month = sep,
	year = {2025},
	pages = {1--19},
}

@article{stoll_professional_2006,
	title = {Professional {Learning} {Communities}: {A} {Review} of the {Literature}},
	volume = {7},
	issn = {1573-1812},
	shorttitle = {Professional {Learning} {Communities}},
	url = {https://doi.org/10.1007/s10833-006-0001-8},
	doi = {10.1007/s10833-006-0001-8},
	language = {en},
	number = {4},
	urldate = {2026-02-06},
	journal = {Journal of Educational Change},
	author = {Stoll, Louise and Bolam, Ray and McMahon, Agnes and Wallace, Mike and Thomas, Sally},
	month = dec,
	year = {2006},
	pages = {221--258},
}

@article{zhang_comparative_2023,
	title = {Comparative interaction patterns of groups in an open network environment: {The} role of facilitators in collaborative learning},
	volume = {40},
	copyright = {© 2023 John Wiley \& Sons Ltd.},
	issn = {1365-2729},
	shorttitle = {Comparative interaction patterns of groups in an open network environment},
	url = {https://onlinelibrary.wiley.com/doi/abs/10.1111/jcal.12873},
	doi = {10.1111/jcal.12873},
	language = {en},
	number = {1},
	urldate = {2023-09-13},
	journal = {Journal of Computer Assisted Learning},
	author = {Zhang, Wenmei and Wang, Cixiao},
	month = sep,
	year = {2023},
	pages = {136--157},
}

@article{bai_how_2025,
	title = {How does peer-led teaching presence promote the development of cognitive presence? {Evidence} based on behavioural sequence analysis},
	volume = {41},
	copyright = {Copyright (c) 2025 xuemei Bai, Xiaoqing Gu},
	issn = {1449-5554, 1449-3098},
	shorttitle = {How does peer-led teaching presence promote the development of cognitive presence?},
	url = {https://ajet.org.au/index.php/AJET/article/view/10020},
	doi = {10.14742/ajet.10020},
	language = {en},
	number = {3},
	urldate = {2025-10-22},
	journal = {Australasian Journal of Educational Technology},
	author = {Bai, Xuemei and Gu, Xiaoqing},
	month = may,
	year = {2025},
	pages = {26--44},
}

@article{kasneci_chatgpt_2023,
	title = {{ChatGPT} for good? {On} opportunities and challenges of large language models for education},
	volume = {103},
	issn = {1041-6080},
	shorttitle = {{ChatGPT} for good?},
	url = {https://www.sciencedirect.com/science/article/pii/S1041608023000195},
	doi = {10.1016/j.lindif.2023.102274},
	urldate = {2023-09-16},
	journal = {Learning and Individual Differences},
	author = {Kasneci, Enkelejda and Sessler, Kathrin and Küchemann, Stefan and Bannert, Maria and Dementieva, Daryna and Fischer, Frank and Gasser, Urs and Groh, Georg and Günnemann, Stephan and Hüllermeier, Eyke and Krusche, Stephan and Kutyniok, Gitta and Michaeli, Tilman and Nerdel, Claudia and Pfeffer, Jürgen and Poquet, Oleksandra and Sailer, Michael and Schmidt, Albrecht and Seidel, Tina and Stadler, Matthias and Weller, Jochen and Kuhn, Jochen and Kasneci, Gjergji},
	month = apr,
	year = {2023},
	pages = {102274},
}

@article{granovetter_strength_1973,
	title = {The {Strength} of {Weak} {Ties}},
	volume = {78},
	issn = {0002-9602},
	url = {https://www.jstor.org/stable/2776392},
	number = {6},
	urldate = {2022-06-01},
	journal = {American Journal of Sociology},
	author = {Granovetter, Mark S.},
	year = {1973},
	pages = {1360--1380},
}

@article{yan_promises_2024,
	title = {Promises and challenges of generative artificial intelligence for human learning},
	volume = {8},
	copyright = {2024 Springer Nature Limited},
	issn = {2397-3374},
	url = {https://www.nature.com/articles/s41562-024-02004-5},
	doi = {10.1038/s41562-024-02004-5},
	language = {en},
	number = {10},
	urldate = {2025-05-31},
	journal = {Nature Human Behaviour},
	author = {Yan, Lixiang and Greiff, Samuel and Teuber, Ziwen and Gašević, Dragan},
	month = oct,
	year = {2024},
	pages = {1839--1850},
}

@article{xu_research_2024,
	title = {Research on the patterns of the organic growth knowledge production based on a dual perspective of content and behaviour},
	volume = {55},
	copyright = {© 2023 British Educational Research Association.},
	issn = {0007-1013, 1467-8535},
	url = {https://bera-journals.onlinelibrary.wiley.com/doi/10.1111/bjet.13418},
	doi = {10.1111/bjet.13418},
	language = {en},
	number = {3},
	urldate = {2024-07-24},
	journal = {British Journal of Educational Technology},
	author = {Xu, Yaqian and Yang, Yang},
	month = may,
	year = {2024},
	pages = {1145--1166},
}

@article{weidlich_extending_2023,
	title = {Extending social presence theory: social presence divergence and interaction integration in online distance learning},
	volume = {35},
	issn = {1042-1726, 1867-1233},
	shorttitle = {Extending social presence theory},
	url = {https://link.springer.com/10.1007/s12528-022-09325-2},
	doi = {10.1007/s12528-022-09325-2},
	language = {en},
	number = {3},
	urldate = {2022-09-07},
	journal = {Journal of Computing in Higher Education},
	author = {Weidlich, Joshua and Göksün, Derya Orhan and Kreijns, Karel},
	month = dec,
	year = {2023},
	pages = {391--412},
}

@article{wang_interaction_2017,
	title = {Interaction pattern analysis in {cMOOCs} based on the connectivist interaction and engagement framework},
	volume = {48},
	issn = {0007-1013, 1467-8535},
	shorttitle = {Interaction pattern analysis in {cMOOCs} based on the connectivist interaction and engagement framework},
	url = {https://bera-journals.onlinelibrary.wiley.com/doi/10.1111/bjet.12433},
	doi = {10.1111/bjet.12433},
	language = {en},
	number = {2},
	urldate = {2021-11-09},
	journal = {British Journal of Educational Technology},
	author = {Wang, Zhijun and Anderson, Terry and Chen, Li and Barbera, Elena},
	month = mar,
	year = {2017},
	pages = {683--699},
}

@article{liu_improving_2024,
	title = {Improving knowledge gain and emotional experience in online learning with knowledge and emotional scaffolding-based conversational agent},
	volume = {27},
	issn = {1176-3647},
	doi = {10.30191/ETS.202404_27(2).RP08},
	language = {en},
	number = {2},
	journal = {Educational Technology \& Society},
	author = {Liu, Zhi and Duan, Huimin and Liu, Shiqi and Mu, Rui and Liu, Sannyuya and Yang, Zongkai},
	month = apr,
	year = {2024},
}

@article{tian_measurement_2026,
	title = {The measurement and characteristic analysis of learner interaction levels in {cMOOCs} based on path analysis},
	volume = {34},
	copyright = {© 2025 Informa UK Limited, trading as Taylor \& Francis Group},
	issn = {1049-4820, 1744-5191},
	url = {https://www.tandfonline.com/doi/full/10.1080/10494820.2025.2498534},
	doi = {10.1080/10494820.2025.2498534},
	language = {EN},
	number = {1},
	urldate = {2025-05-08},
	journal = {Interactive Learning Environments},
	author = {Tian, Yulin and Xiao, Jianjun},
	month = jan,
	year = {2026},
	pages = {297--319},
}

@article{song_interactions_2025,
	title = {Interactions with generative {AI} chatbots: unveiling dialogic dynamics, students’ perceptions, and practical competencies in creative problem-solving},
	volume = {22},
	issn = {2365-9440},
	shorttitle = {Interactions with generative {AI} chatbots},
	url = {https://educationaltechnologyjournal.springeropen.com/articles/10.1186/s41239-025-00508-2},
	doi = {10.1186/s41239-025-00508-2},
	language = {en-US},
	number = {1},
	urldate = {2025-03-06},
	journal = {International Journal of Educational Technology in Higher Education},
	author = {Song, Yu and Huang, Longchao and Zheng, Lanqin and Fan, Mengya and Liu, Zehao},
	month = mar,
	year = {2025},
	pages = {12},
}

@article{siemens_connectivism_2005,
	title = {Connectivism: {A} learning theory for the digital age},
	volume = {2},
	url = {https://www.scopus.com/inward/record.uri?eid=2-s2.0-33745587105&partnerID=40&md5=63cd1564999432bc8cc6c44bcbde840e},
	language = {undefined},
	number = {1},
	journal = {International Journal of Instructional Technology and Distance Learning},
	author = {Siemens, G.},
	year = {2005},
	note = {tex.publication\_stage: Final},
}

@article{shea_reexamination_2010,
	series = {Special {Issue} on the {Community} of {Inquiry} {Framework}: {Ten} {Years} {Later}},
	title = {A re-examination of the community of inquiry framework: {Social} network and content analysis},
	volume = {13},
	issn = {10967516},
	shorttitle = {A re-examination of the community of inquiry framework},
	url = {https://linkinghub.elsevier.com/retrieve/pii/S1096751609000682},
	doi = {10.1016/j.iheduc.2009.11.002},
	language = {en},
	number = {1-2},
	urldate = {2022-08-30},
	journal = {The Internet and Higher Education},
	author = {Shea, Peter and Hayes, Suzanne and Vickers, Jason and Gozza-Cohen, Mary and Uzuner, Sedef and Mehta, Ruchi and Valchova, Anna and Rangan, Prahalad},
	month = jan,
	year = {2010},
	pages = {10--21},
}

@article{wang_what_2023,
	title = {What matters in {AI}-supported learning: {A} study of human-{AI} interactions in language learning using cluster analysis and epistemic network analysis},
	volume = {194},
	issn = {03601315},
	shorttitle = {What matters in {AI}-supported learning},
	url = {https://linkinghub.elsevier.com/retrieve/pii/S0360131522002743},
	doi = {10.1016/j.compedu.2022.104703},
	language = {en},
	urldate = {2022-12-09},
	journal = {Computers \& Education},
	author = {Wang, Xinghua and Liu, Qian and Pang, Hui and Tan, Seng Chee and Lei, Jun and Wallace, Matthew P. and Li, Linlin},
	month = mar,
	year = {2023},
	pages = {104703},
}

@article{wang_who_2024,
	title = {Who will participate in online collaborative problem solving? {A} longitudinal network analysis},
	volume = {32},
	issn = {1049-4820, 1744-5191},
	shorttitle = {Who will participate in online collaborative problem solving?},
	url = {https://www.tandfonline.com/doi/full/10.1080/10494820.2023.2266523},
	doi = {10.1080/10494820.2023.2266523},
	language = {en-US},
	number = {10},
	urldate = {2023-10-09},
	journal = {Interactive Learning Environments},
	author = {Wang, Cixiao and Xiao, Jianjun},
	month = nov,
	year = {2024},
	pages = {6534--6551},
}

@article{wise_learning_2018,
	title = {Learning communities in the crowd: {Characteristics} of content related interactions and social relationships in {MOOC} discussion forums},
	volume = {122},
	issn = {03601315},
	shorttitle = {Learning communities in the crowd},
	url = {https://linkinghub.elsevier.com/retrieve/pii/S0360131518300800},
	doi = {10.1016/j.compedu.2018.03.021},
	language = {en},
	urldate = {2023-07-28},
	journal = {Computers \& Education},
	author = {Wise, Alyssa Friend and Cui, Yi},
	month = jul,
	year = {2018},
	pages = {221--242},
}

@article{shea_building_2022,
	title = {Building bridges to advance the {Community} of {Inquiry} framework for online learning},
	volume = {57},
	issn = {0046-1520, 1532-6985},
	url = {https://www.tandfonline.com/doi/full/10.1080/00461520.2022.2089989},
	doi = {10.1080/00461520.2022.2089989},
	language = {en-US},
	number = {3},
	urldate = {2024-01-24},
	journal = {Educational Psychologist},
	author = {Shea, Peter and Richardson, Jennifer and Swan, Karen},
	month = jul,
	year = {2022},
	pages = {148--161},
}

@article{rolim_networkbased_2019,
	title = {A network-based analytic approach to uncovering the relationship between social and cognitive presences in communities of inquiry},
	volume = {42},
	issn = {10967516},
	url = {https://linkinghub.elsevier.com/retrieve/pii/S1096751619300235},
	doi = {10.1016/j.iheduc.2019.05.001},
	language = {en-US},
	urldate = {2023-09-28},
	journal = {The Internet and Higher Education},
	author = {Rolim, Vitor and Ferreira, Rafael and Lins, Rafael Dueire and Gǎsević, Dragan},
	month = jul,
	year = {2019},
	pages = {53--65},
}

@article{nguyen_role_2023,
	title = {Role design considerations of conversational agents to facilitate discussion and systems thinking},
	volume = {192},
	issn = {03601315},
	url = {https://linkinghub.elsevier.com/retrieve/pii/S0360131522002329},
	doi = {10.1016/j.compedu.2022.104661},
	language = {en-US},
	urldate = {2024-01-18},
	journal = {Computers \& Education},
	author = {Nguyen, Ha},
	month = jan,
	year = {2023},
	pages = {104661},
}

@article{prestridge_identifying_2024,
	title = {Identifying how classroom teachers develop presence online: breaking the fourth wall in online learning},
	volume = {29},
	issn = {1360-2357, 1573-7608},
	shorttitle = {Identifying how classroom teachers develop presence online},
	url = {https://link.springer.com/10.1007/s10639-023-11714-8},
	doi = {10.1007/s10639-023-11714-8},
	language = {en},
	number = {2},
	urldate = {2023-03-23},
	journal = {Education and Information Technologies},
	author = {Prestridge, Sarah and Main, Katherine and Schmid, Mirjam},
	month = feb,
	year = {2024},
	pages = {1357--1377},
}

@article{ortega-ochoa_empathic_2024,
	title = {Empathic pedagogical conversational agents: {A} systematic literature review},
	volume = {55},
	copyright = {© 2023 The Authors. British Journal of Educational Technology published by John Wiley \& Sons Ltd on behalf of British Educational Research Association.},
	issn = {0007-1013, 1467-8535},
	shorttitle = {Empathic pedagogical conversational agents},
	url = {https://bera-journals.onlinelibrary.wiley.com/doi/10.1111/bjet.13413},
	doi = {10.1111/bjet.13413},
	language = {en},
	number = {3},
	urldate = {2023-12-08},
	journal = {British Journal of Educational Technology},
	author = {Ortega‐Ochoa, Elvis and Arguedas, Marta and Daradoumis, Thanasis},
	month = may,
	year = {2024},
	pages = {886--909},
}

@inproceedings{natarajan_humanintheloop_2025,
	title = {Human-in-the-loop or {AI}-in-the-loop? {Automate} or {Collaborate}?},
	volume = {39},
	shorttitle = {Human-in-the-loop or {AI}-in-the-loop?},
	url = {https://ojs.aaai.org/index.php/AAAI/article/view/35083},
	number = {27},
	urldate = {2025-06-16},
	booktitle = {Proceedings of the {AAAI} {Conference} on {Artificial} {Intelligence}},
	author = {Natarajan, Sriraam and Mathur, Saurabh and Sidheekh, Sahil and Stammer, Wolfgang and Kersting, Kristian},
	year = {2025},
	pages = {28594--28600},
}

@article{li_dynamics_2025,
	title = {Dynamics of {cMOOC} learner interactions in different social media: a longitudinal network analysis},
	volume = {33},
	issn = {1049-4820, 1744-5191},
	shorttitle = {Dynamics of {cMOOC} learner interactions in different social media},
	url = {https://www.tandfonline.com/doi/full/10.1080/10494820.2024.2412058},
	doi = {10.1080/10494820.2024.2412058},
	language = {en-US},
	number = {3},
	urldate = {2024-10-09},
	journal = {Interactive Learning Environments},
	author = {Li, Shuang and Du, Junlei and Yu, Xinpei},
	month = mar,
	year = {2025},
	pages = {2460--2478},
}

@article{liu_looking_2022,
	title = {Looking at {MOOC} discussion data to uncover the relationship between discussion pacings, learners’ cognitive presence and learning achievements},
	volume = {27},
	copyright = {2022 The Author(s), under exclusive licence to Springer Science+Business Media, LLC, part of Springer Nature},
	issn = {1360-2357, 1573-7608},
	url = {https://link.springer.com/10.1007/s10639-022-10943-7},
	doi = {10.1007/s10639-022-10943-7},
	language = {en},
	number = {6},
	urldate = {2022-10-03},
	journal = {Education and Information Technologies},
	author = {Liu, Zhi and Kong, Xi and Liu, Sannyuya and Yang, Zongkai and Zhang, Cuishuang},
	month = jul,
	year = {2022},
	pages = {8265--8288},
}

@article{li_exploring_2022,
	title = {Exploring the relationship between interaction patterns and social capital accumulation in connectivist learning},
	volume = {0},
	issn = {1049-4820, 1744-5191},
	url = {https://www.tandfonline.com/doi/full/10.1080/10494820.2022.2157839},
	doi = {10.1080/10494820.2022.2157839},
	language = {en-US},
	number = {0},
	urldate = {2022-12-23},
	journal = {Interactive Learning Environments},
	author = {Li, Shuang and He, Xinyi and Chen, Jiaqi},
	month = dec,
	year = {2022},
	pages = {1--22},
}

@article{kop_pedagogy_2011,
	title = {A pedagogy of abundance or a pedagogy to support human beings? {Participant} support on massive open online courses},
	volume = {12},
	shorttitle = {A pedagogy of abundance or a pedagogy to support human beings?},
	number = {7},
	journal = {International Review of Research in Open and Distributed Learning},
	author = {Kop, Rita and Fournier, Hélène and Mak, John Sui Fai},
	year = {2011},
	pages = {74--93},
}

@article{johnson_pedagogical_2018,
	title = {Pedagogical {Agents}: {Back} to the {Future}},
	volume = {39},
	copyright = {© 2018 The Authors. AI Magazine published by John Wiley \& Sons Ltd on behalf of Association for the Advancement of Artificial Intelligence},
	issn = {0738-4602, 2371-9621},
	shorttitle = {Pedagogical {Agents}},
	url = {https://onlinelibrary.wiley.com/doi/10.1609/aimag.v39i2.2793},
	doi = {10.1609/aimag.v39i2.2793},
	language = {en},
	number = {2},
	urldate = {2024-01-19},
	journal = {AI Magazine},
	author = {Johnson, W. Lewis and Lester, James C.},
	month = jun,
	year = {2018},
	pages = {33--44},
}

@article{huang_facilitating_2025,
	title = {Facilitating {Online} {Self}-{Regulated} {Learning} and {Social} {Presence} {Using} {Chatbots}: {Evidence}-{Based} {Design} {Principles}},
	volume = {18},
	issn = {1939-1382, 2372-0050},
	shorttitle = {Facilitating {Online} {Self}-{Regulated} {Learning} and {Social} {Presence} {Using} {Chatbots}},
	url = {https://ieeexplore.ieee.org/document/10816550/},
	doi = {10.1109/TLT.2024.3523199},
	language = {en-US},
	urldate = {2025-01-19},
	journal = {IEEE Transactions on Learning Technologies},
	author = {Huang, Weijiao and Hew, Khe Foon},
	year = {2025},
	pages = {56--71},
}

@article{kong_exploratory_2025,
	title = {Exploratory study of an {AI}-supported discussion representational tool for online collaborative learning in a {Chinese} university},
	volume = {64},
	issn = {10967516},
	url = {https://linkinghub.elsevier.com/retrieve/pii/S1096751624000356},
	doi = {10.1016/j.iheduc.2024.100973},
	urldate = {2025-01-13},
	journal = {The Internet and Higher Education},
	author = {Kong, Xi and Liu, Zhi and Chen, Changsheng and Liu, Sannyuya and Xu, Zhenguo and Tang, Qianhui},
	month = jan,
	year = {2025},
	pages = {100973},
}

@article{knox_learning_2018,
	title = {Beyond the “c” and the “x”: {Learning} with algorithms in massive open online courses ({MOOCs})},
	volume = {64},
	issn = {0020-8566, 1573-0638},
	shorttitle = {Beyond the “c” and the “x”},
	url = {http://link.springer.com/10.1007/s11159-018-9707-0},
	doi = {10.1007/s11159-018-9707-0},
	language = {en},
	number = {2},
	urldate = {2025-06-01},
	journal = {International Review of Education},
	author = {Knox, Jeremy},
	month = apr,
	year = {2018},
	pages = {161--178},
}

@article{joksimovic_exploring_2018,
	title = {Exploring development of social capital in a {CMOOC} through language and discourse},
	volume = {36},
	issn = {10967516},
	shorttitle = {🔤从语言与话语看中国大型网络公开课中社会资本的发展},
	url = {https://linkinghub.elsevier.com/retrieve/pii/S1096751617304554},
	doi = {10.1016/j.iheduc.2017.09.004},
	language = {en},
	urldate = {2021-11-09},
	journal = {The Internet and Higher Education},
	author = {Joksimović, Srećko and Dowell, Nia and Poquet, Oleksandra and Kovanović, Vitomir and Gašević, Dragan and Dawson, Shane and Graesser, Arthur C.},
	month = jan,
	year = {2018},
	pages = {54--64},
}

@article{huang_chatbots_2022,
	title = {Chatbots for language learning—{Are} they really useful? {A} systematic review of chatbot‐supported language learning},
	volume = {38},
	copyright = {© 2021 John Wiley \& Sons Ltd},
	issn = {0266-4909, 1365-2729},
	shorttitle = {Chatbots for language learning—{Are} they really useful?},
	url = {https://onlinelibrary.wiley.com/doi/10.1111/jcal.12610},
	doi = {10.1111/jcal.12610},
	language = {en},
	number = {1},
	urldate = {2024-01-18},
	journal = {Journal of Computer Assisted Learning},
	author = {Huang, Weijiao and Hew, Khe Foon and Fryer, Luke K.},
	month = feb,
	year = {2022},
	pages = {237--257},
}

@article{songheehan_equity_2024,
	title = {Equity at the forefront: {A} systematic research and development process of chatbot curriculum for massive open online courses},
	volume = {27},
	issn = {1176-3647},
	doi = {10.30191/ETS.202410_27(4).SP08},
	number = {4},
	journal = {Educational Technology \& Society},
	author = {{Songhee Han} and {Min Liu}},
	month = oct,
	year = {2024},
	pages = {339--351},
}

@article{han_faq_2022,
	title = {{FAQ} chatbot and inclusive learning in massive open online courses},
	volume = {179},
	url = {https://www.sciencedirect.com/science/article/pii/S0360131521002724},
	urldate = {2025-01-17},
	journal = {Computers \& Education},
	author = {Han, Songhee and Lee, Min Kyung},
	year = {2022},
	pages = {104395},
}

@article{guo_community_2021,
	title = {The {Community} of {Inquiry} perspective on students' social presence, cognitive presence, and academic performance in online project‐based learning},
	volume = {37},
	copyright = {© 2021 The Authors. Journal of Computer Assisted Learning published by John Wiley \& Sons Ltd.},
	issn = {0266-4909, 1365-2729},
	url = {https://onlinelibrary.wiley.com/doi/10.1111/jcal.12586},
	doi = {10.1111/jcal.12586},
	language = {en},
	number = {5},
	urldate = {2023-09-16},
	journal = {Journal of Computer Assisted Learning},
	author = {Guo, Pengyue and Saab, Nadira and Wu, Lin and Admiraal, Wilfried},
	month = oct,
	year = {2021},
	pages = {1479--1493},
}

@article{garrison_critical_2001,
	title = {Critical thinking, cognitive presence, and computer conferencing in distance education},
	volume = {15},
	issn = {0892-3647, 1538-9286},
	url = {http://www.tandfonline.com/doi/abs/10.1080/08923640109527071},
	doi = {10.1080/08923640109527071},
	language = {en-US},
	number = {1},
	urldate = {2022-09-17},
	journal = {American Journal of Distance Education},
	author = {Garrison, D. Randy and Anderson, Terry and Archer, Walter},
	month = jan,
	year = {2001},
	pages = {7--23},
}

@article{garrison_critical_1999,
	title = {Critical {Inquiry} in a {Text}-{Based} {Environment}: {Computer} {Conferencing} in {Higher} {Education}},
	volume = {2},
	issn = {10967516},
	shorttitle = {Critical {Inquiry} in a {Text}-{Based} {Environment}},
	url = {https://linkinghub.elsevier.com/retrieve/pii/S1096751600000166},
	doi = {10.1016/S1096-7516(00)00016-6},
	language = {en},
	number = {2-3},
	urldate = {2022-06-01},
	journal = {The Internet and Higher Education},
	author = {Garrison, D.Randy and Anderson, Terry and Archer, Walter},
	month = mar,
	year = {1999},
	pages = {87--105},
}

@article{fuchs_exploring_2023,
	title = {Exploring the opportunities and challenges of {NLP} models in higher education: is {Chat} {GPT} a blessing or a curse?},
	volume = {8},
	issn = {2504-284X},
	shorttitle = {Exploring the opportunities and challenges of {NLP} models in higher education},
	url = {https://www.frontiersin.org/articles/10.3389/feduc.2023.1166682},
	urldate = {2023-09-14},
	journal = {Frontiers in Education},
	author = {Fuchs, Kevin},
	year = {2023},
}

@article{fidalgo-blanco_massive_2016,
	title = {From massive access to cooperation: lessons learned and proven results of a hybrid {xMOOC}/{cMOOC} pedagogical approach to {MOOCs}},
	volume = {13},
	issn = {2365-9440},
	shorttitle = {From massive access to cooperation},
	url = {http://educationaltechnologyjournal.springeropen.com/articles/10.1186/s41239-016-0024-z},
	doi = {10.1186/s41239-016-0024-z},
	language = {en},
	number = {1},
	urldate = {2025-06-01},
	journal = {International Journal of Educational Technology in Higher Education},
	author = {Fidalgo-Blanco, Ángel and Sein-Echaluce, María Luisa and García-Peñalvo, Francisco José},
	month = dec,
	year = {2016},
	pages = {24},
}

@article{castellanos-reyes_dynamics_2021,
	title = {The dynamics of a {MOOC}'s learner-learner interaction over time: {A} longitudinal network analysis},
	volume = {123},
	issn = {07475632},
	shorttitle = {🔤大型开放式网络课程中学习者与学习者互动的动态过程: 纵向网络分析},
	url = {https://linkinghub.elsevier.com/retrieve/pii/S074756322100203X},
	doi = {10.1016/j.chb.2021.106880},
	language = {en},
	urldate = {2021-11-09},
	journal = {Computers in Human Behavior},
	author = {Castellanos-Reyes, Daniela},
	month = oct,
	year = {2021},
	pages = {106880},
}

@article{chen_theoretical_2022,
	title = {Theoretical {Development} of {Connectivism} through {Innovative} {Application} in {China}},
	volume = {48},
	copyright = {Copyright (c) 2022 Li Chen, Yaqian Xu},
	issn = {1499-6685, 1499-6677},
	url = {https://cjlt.ca/index.php/cjlt/article/view/28255},
	doi = {10.21432/cjlt28255},
	language = {en},
	number = {4},
	urldate = {2022-12-30},
	journal = {Canadian Journal of Learning and Technology},
	author = {Chen, Li and Xu, Yaqian},
	month = nov,
	year = {2022},
	note = {Number: 4},
}

@article{bakki_moocat_2019,
	title = {{MOOCAT}: {A} visual authoring tool in the {cMOOC} context},
	volume = {24},
	copyright = {2018 Springer Science+Business Media, LLC, part of Springer Nature},
	issn = {1360-2357, 1573-7608},
	shorttitle = {{MOOCAT}},
	url = {http://link.springer.com/10.1007/s10639-018-9807-2},
	doi = {10.1007/s10639-018-9807-2},
	language = {en},
	number = {2},
	urldate = {2023-04-08},
	journal = {Education and Information Technologies},
	author = {Bakki, Aicha and Oubahssi, Lahcen and George, Sébastien and Cherkaoui, Chihab},
	month = mar,
	year = {2019},
	pages = {1185--1209},
}

@article{bai_impact_2023,
	title = {The impact of {cMOOC} learners’ interaction on content production},
	volume = {31},
	copyright = {All rights reserved},
	issn = {1049-4820, 1744-5191},
	url = {https://www.tandfonline.com/doi/full/10.1080/10494820.2021.1969955},
	doi = {10.1080/10494820.2021.1969955},
	language = {en},
	number = {7},
	urldate = {2021-11-09},
	journal = {Interactive Learning Environments},
	author = {Bai, Yun-Qi and Xiao, Jian-Jun},
	month = oct,
	year = {2023},
	note = {15 citations (CrossRef 2026/1/31)},
	pages = {4464--4475},
}

@article{ba_investigating_2025,
	title = {Investigating the impact of {\textless}span style="font-variant:small-caps;"{\textgreater}{ChatGPT}{\textless}/span{\textgreater} ‐assisted feedback on the dynamics and outcomes of online inquiry‐based discussion},
	volume = {56},
	copyright = {© 2025 British Educational Research Association.},
	issn = {0007-1013, 1467-8535},
	url = {https://bera-journals.onlinelibrary.wiley.com/doi/10.1111/bjet.13605},
	doi = {10.1111/bjet.13605},
	language = {en},
	number = {5},
	urldate = {2025-05-21},
	journal = {British Journal of Educational Technology},
	author = {Ba, Shen and Zhan, Ying and Huang, Lingyun and Lu, Guoqing},
	month = sep,
	year = {2025},
	pages = {1710--1734},
}

@article{anderson_three_2011,
	title = {Three generations of distance education pedagogy},
	volume = {12},
	issn = {1492-3831},
	url = {http://www.irrodl.org/index.php/irrodl/article/view/890},
	doi = {10.19173/irrodl.v12i3.890},
	language = {en},
	number = {3},
	urldate = {2022-05-17},
	journal = {The International Review of Research in Open and Distributed Learning},
	author = {Anderson, Terry and Dron, Jon},
	month = mar,
	year = {2011},
	pages = {80},
}

@article{bai_more_2023,
	title = {More factors, better understanding: model verification and construct validity study on the community of inquiry in {MOOC}},
	volume = {28},
	issn = {1360-2357, 1573-7608},
	shorttitle = {More factors, better understanding},
	url = {https://link.springer.com/10.1007/s10639-023-11604-z},
	doi = {10.1007/s10639-023-11604-z},
	language = {en},
	number = {8},
	urldate = {2023-09-16},
	journal = {Education and Information Technologies},
	author = {Bai, Xuemei and Gu, Xiaoqing and Guo, Rifa},
	month = aug,
	year = {2023},
	pages = {10483--10506},
}

@article{kovanovic_exploring_2018,
	title = {Exploring communities of inquiry in {Massive} {Open} {Online} {Courses}},
	volume = {119},
	issn = {03601315},
	url = {https://linkinghub.elsevier.com/retrieve/pii/S0360131517302543},
	doi = {10.1016/j.compedu.2017.11.010},
	language = {en-US},
	urldate = {2023-09-16},
	journal = {Computers \& Education},
	author = {Kovanović, Vitomir and Joksimović, Srećko and Poquet, Oleksandra and Hennis, Thieme and Čukić, Iva and De Vries, Pieter and Hatala, Marek and Dawson, Shane and Siemens, George and Gašević, Dragan},
	month = apr,
	year = {2018},
	pages = {44--58},
}

@article{akyol_impact_2011,
	title = {The impact of course duration on the development of a community of inquiry},
	volume = {19},
	issn = {1049-4820, 1744-5191},
	url = {http://www.tandfonline.com/doi/abs/10.1080/10494820902809147},
	doi = {10.1080/10494820902809147},
	language = {en-US},
	number = {3},
	urldate = {2024-01-17},
	journal = {Interactive Learning Environments},
	author = {Akyol, Zehra and Vaughan, Norm and Garrison, D. Randy},
	month = jun,
	year = {2011},
	pages = {231--246},
}

@article{kozan_interrelationships_2014,
	title = {Interrelationships between and among social, teaching, and cognitive presence},
	volume = {21},
	issn = {10967516},
	url = {https://linkinghub.elsevier.com/retrieve/pii/S1096751613000511},
	doi = {10.1016/j.iheduc.2013.10.007},
	urldate = {2023-09-28},
	journal = {The Internet and Higher Education},
	author = {Kozan, Kadir and Richardson, Jennifer C.},
	month = apr,
	year = {2014},
	pages = {68--73},
}

@article{zou_exploring_2021,
	title = {Exploring the relationship between social presence and learners’ prestige in {MOOC} discussion forums using automated content analysis and social network analysis},
	volume = {115},
	issn = {07475632},
	url = {https://linkinghub.elsevier.com/retrieve/pii/S0747563220303290},
	doi = {10.1016/j.chb.2020.106582},
	language = {en},
	urldate = {2023-08-01},
	journal = {Computers in Human Behavior},
	author = {Zou, Wenting and Hu, Xiao and Pan, Zilong and Li, Chenglu and Cai, Ying and Liu, Min},
	month = feb,
	year = {2021},
	pages = {106582},
}

@article{poquet_social_2018,
	title = {Social presence in massive open online courses},
	volume = {19},
	language = {en-US},
	number = {3},
	journal = {International Review of Research in Open and Distributed Learning},
	author = {Poquet, Oleksandra and Kovanović, Vitomir and de Vries, Pieter and Hennis, Thieme and Joksimović, Srećko and Gašević, Dragan and Dawson, Shane},
	year = {2018},
}

@article{richardson_social_2017,
	title = {Social presence in relation to students' satisfaction and learning in the online environment: {A} meta-analysis},
	volume = {71},
	issn = {07475632},
	shorttitle = {Social presence in relation to students' satisfaction and learning in the online environment},
	url = {https://linkinghub.elsevier.com/retrieve/pii/S0747563217300742},
	doi = {10.1016/j.chb.2017.02.001},
	language = {en-US},
	urldate = {2023-08-30},
	journal = {Computers in Human Behavior},
	author = {Richardson, Jennifer C. and Maeda, Yukiko and Lv, Jing and Caskurlu, Secil},
	month = jun,
	year = {2017},
	pages = {402--417},
}

@article{feng_group_2025,
	title = {Group interaction patterns in generative {\textless}span style="font-variant:small-caps;"{\textgreater}{AI}{\textless}/span{\textgreater} ‐supported collaborative problem solving: {Network} analysis of the interactions among students and a {\textless}span style="font-variant:small-caps;"{\textgreater}{GAI}{\textless}/span{\textgreater} chatbot},
	volume = {56},
	copyright = {© 2025 The Author(s). British Journal of Educational Technology published by John Wiley \& Sons Ltd on behalf of British Educational Research Association.},
	issn = {0007-1013, 1467-8535},
	shorttitle = {Group interaction patterns in generative {AI}-supported collaborative problem solving},
	url = {https://bera-journals.onlinelibrary.wiley.com/doi/10.1111/bjet.13611},
	doi = {10.1111/bjet.13611},
	language = {en},
	number = {5},
	urldate = {2025-06-30},
	journal = {British Journal of Educational Technology},
	author = {Feng, Shihui},
	month = sep,
	year = {2025},
	pages = {2125--2145},
}


\end{document}